\documentclass[prb,amsmath,amssymb,twocolumn,showpacs,floatfix,superscriptaddress]{revtex4}
\usepackage{graphicx}
\usepackage[english]{babel}
\usepackage{times}
\usepackage{dcolumn}
\usepackage[usenames,dvipsnames]{color}
\usepackage{units}

\begin{document}

\title{SU(3) Kondo effect in spinless triple quantum dots}

\author{Rosa L\'{o}pez}
 \affiliation{Institut de F\'{\i}sica Interdisciplin\`aria i de Sistemes Complexos
IFISC (CSIC-UIB), E-07122 Palma de Mallorca, Spain}
\affiliation{Departament de F\'{i}sica, Universitat de les Illes Balears,
  E-07122 Palma de Mallorca, Spain}

\author{Toma\v{z} Rejec}
\affiliation{Faculty  of Mathematics and Physics, University of Ljubljana,
Jadranska 19, SI-1000 Ljubljana, Slovenia}
\affiliation{Jo\v{z}ef Stefan Institute, Jamova 39, SI-1000 Ljubljana,
Slovenia}

\author{Jan Martinek}
\affiliation{Institute of Molecular Physics,
 Polish Academy of Sciences, Smoluchowskiego 17,
 60-179 Pozna\'n, Poland}

\author{Rok \v{Z}itko}
\affiliation{Jo\v{z}ef Stefan Institute, Jamova 39, SI-1000 Ljubljana,
Slovenia}
\affiliation{Faculty  of Mathematics and Physics, University of Ljubljana,
Jadranska 19, SI-1000 Ljubljana, Slovenia}

\date{\today}

 
% 72.10.Fk Scattering by point defects, dislocations, surfaces, and
% other imperfections (including Kondo effect)
% 72.15.Qm Scattering mechanisms and Kondo effect (see also 75.20.Hr
% Local moments in compounds and alloys; Kondo effect, valence
% fluctuations, heavy fermions in magnetic properties and materials)
% 72.25.Dc Spin polarized transport in semiconductors 
% 73.20.-r Electron states at surfaces and interfaces
% 73.20.Hb Impurity and defect levels; energy states of adsorbed species 

\begin{abstract}
We discuss a device --- a purely capacitively coupled interacting
spinless triple quantum dot system --- for the observation of the
SU(3) Kondo effect. Unlike more familiar SU(2) and SU(4) Kondo effects
in quantum dot devices which lead to unitary linear conductance at low
temperatures, the SU(3) Kondo scenario can be easily identified by the
conductance pinned to a characteristic value of $3/4$ of the unitary
limit. This is associated with the interesting fact that the SU(3)
Kondo effect does not occur at the particle-hole symmetric point,
where the system is found instead in the valence-fluctuating regime
with the total dot occupancy flipping between 1 and 2, but for gate
voltages in the two Kondo plateaux where the dot occupancy is pinned
to an integer value, either 1 or 2. From the thermodynamic analysis in
the Kondo regime we find that the effective impurity orbital moment,
defined through the impurity orbital susceptibility
($\chi_\mathrm{imp}$) multiplied by the temperature, is
$T\chi_\mathrm{imp}=1$ at high temperatures and then it increases to
the characteristic value of $T\chi_\mathrm{imp}=4/3$ corresponding to
the three-fold degenerate local-moment fixed point where the impurity
entropy is $S_\mathrm{imp}=\ln 3$. Then, at much lower temperatures,
the system flows to the non-degenerate strong-coupling fixed point in
which the SU(3) Kondo effect takes place. We also report results about
the robustness of the SU(3) Kondo effect against various perturbations
present in real experimental setups, namely, unequal reservoir-dot
tunneling couplings, gating effects and non-vanishing interdot
tunneling rates. Finally, we describe possible mechanisms to restore
the SU(3) Kondo physics by properly tuning the on-site dot potentials.
We briefly comment on the spinfull case which has very different
behavior and shows Kondo plateaus in conductance for all integer
values of the occupancy, including at the particle-hole symmetric
point.
\end{abstract}

\maketitle

\newcommand{\vc}[1]{{\mathbf{#1}}}
\newcommand{\vck}{\vc{k}}
\newcommand{\braket}[2]{\langle#1|#2\rangle}
\newcommand{\expv}[1]{\langle #1 \rangle}
\newcommand{\ket}[1]{| #1 \rangle}
\newcommand{\Tr}{\mathrm{Tr}}

\newcommand{\SU}{\mathrm{SU}}

\section{Introduction} In metals, magnetic impurities are responsible
for the anomalous behavior of the resistivity at low temperatures.
\cite{hewson1993} Magnetic interactions result from high-order
correlated tunneling events of electrons that are hopping \textit{in}
and \textit{out} of a localized impurity. In this fashion, the
impurity spin is screened through the formation of the Kondo spin
singlet state. In semiconductor quantum dots, the Kondo effect can
also take place as theoretically predicted 
\cite{ng1988,glazman1988,meir1993} and experimentally observed.
\cite{goldhabergordon1998b,cronenwett1998,wiel2000} However, whereas
the resistivity of a magnetically doped metal increases when the
temperature is lowered below the Kondo temperature, $T_K$, in quantum
dots (QDs) the linear conductance increases and eventually reaches its
maximum value $G=2e^2/h$ at zero temperature. \cite{wiel2000} This is
due to the enhancement of the scattering rate which results in the
opening of a channel for electrons that are perfectly transmitted
through the QD. The main advantage of exploring the spin-$1/2$ Kondo
effect in manufactured nanostructures such as QDs
\cite{kouwenhoven1997,goldhabergordon1998b,cronenwett1998,Schmid1998182}
is their high tunability. Using gate electrodes it is possible to vary
in a controllable manner the number of trapped electrons and the
strength of tunnel coupling between the reservoirs and the localized
dot states. Furthermore, as an additional advantage, we mention that
QDs constitute perfect laboratories to test many-body effects under
non-equilibrium conditions.
\cite{ralph1994equib,Kogan28052004,PhysRevLett.89.156801,PhysRevB.71.035315,aguado2000}
The influence of external fields such as finite bias voltages,
\cite{PhysRevLett.89.156801,PhysRevB.71.035315} or time varying ac
fields \cite{Kogan28052004} allows the observation of the Kondo effect
out of equilibrium. Remarkably, magnetic fields dramatically affect
the Kondo state, even at equilibrium conditions.  The level
degeneracy, required for the formation of the usual spin-$1/2$ Kondo
effect, is lifted in the presence of a magnetic field.  There are,
however, some exceptions where a magnetic field facilitates the
development of a Kondo state, for example in vertical QDs 
with an even number of electrons. Here, the integer-spin Kondo effect 
occurs when the singlet and triplet states become degenerate because
of the presence of the magnetic field.
\cite{fuhrer2004b,hofstetter2004,
kogan2003,pustilnik2003,pustilnik2001st,sasaki2000} Therefore, in
general, magnetic fields either remove the Kondo effect or facilitate
its observation through the level degeneracy requirement.

\begin{figure}[htbp]
\centering 
\includegraphics[clip,width=5cm, angle=90]{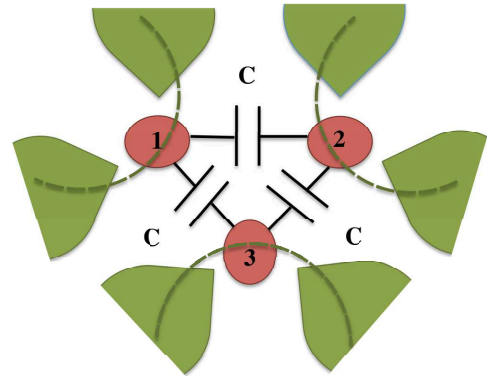}
\caption{(Color online) Schematic representation of the capacitively
coupled triple quantum dot system. Each quantum dot is attached to two
electron reservoirs. We assume that a sufficiently large external
magnetic field is applied to fully polarize the electrons, so that we
may consider the system to be spinless. The only interaction between
each pair of quantum dots is purely capacitive. Inter-dot charging
energies (denoted by $V$) are assumed to be the same: they are
characterized by the capacitance $C$, $V=e^2 /2C$.  Dashed lines
indicate the electron transport through each dot.} \label{schematic}
\end{figure}

The fabrication of more complex nanostructures has opened new
possibilities for the study of unconventional Kondo effect.  There
exists a great variety of artificially fabricated systems exhibiting
exotic kinds of the Kondo effect: to mention just a few, nanodevices
based on carbon materials like fullerenes, carbon nanotubes, and
graphene. \cite{nygard2000,odom2000,jarillo,PhysRevB.74.205119,
PhysRevLett.99.066801,PhysRevB.82.085423,PhysRevB.83.125404,
PhysRevLett.102.046801,PhysRevLett.100.086809,PhysRevB.83.155325}
Others systems, such as $p$-doped nanowires, have demonstrated the
Kondo effect assisted by holes.  \cite{PhysRevLett.107.076805} The
search for highly symmetric Kondo singlets has been revived in carbon
nanotubes and vertical double dot systems with the experimental
demonstration of the $\SU(4)$ Kondo effect.
\cite{PhysRevLett.90.026602,
PhysRevLett.99.066801,jarillo,PhysRevB.71.115312,PhysRevLett.93.017205}
In carbon nanotubes, the valley isospin together with the spin degree
of freedom manifests as a four-fold shell structure in the Coulomb
blockade regime.  \cite{PhysRevLett.88.126801,PhysRevLett.89.046803}
In the low-temperature regime the fluctuations among the four quantum
states lead to the observation of the $\SU(4)$ Kondo effect.
\cite{jarillo,PhysRevLett.95.067204,goldhaber2012} So far, the
$\SU(2)$ and $\SU(4)$ Kondo effects have been extensively studied.
There is, however, very few works devoted to other possible symmetries
for a Kondo singlet, and in particular to the $\SU(3)$ Kondo effect.
In a recent work \cite{PhysRevLett.106.106401} the $\SU(3)$ Kondo
effect has been suggested to be observable in triple quantum dot (TQD)
in the quantum Hall regime.  Here, we discuss a different setup --- a
spinless TQD with equal capacitative coupling $V_i$ between all
quantum dot pairs, as shown in the sketch in Fig.~\ref{schematic}, as
a suitable system for the observation of the $\SU(3)$ Kondo effect.
The $\SU(3)$ Kondo physics takes place when there is a single electron
or a single hole in the whole system, i.e., when either $n=1$ (one
electron and two holes) or $n=2$ (two electrons and one hole).  This
defines three possible {\it flavors} corresponding to the position of
the electron (or hole) in one of the three dots or leads.  In this
work, we will refer to the flavor degree of freedom also as the {\it
channel} or {\it orbital} degree of freedom; for our spinless model,
these expressions are fully interchangeable. Each dot is connected to
two contacts in such a way that the tunneling events conserve the
flavor degree of freedom.  Notice that in principle, this setup can
be easily generalized to build an arbitrary $\SU(N)$ Kondo state 
\cite{PhysRevB.80.125304,PhysRevB.80.155322,PhysRevB.83.241301,
PhysRevLett.106.106401,PhysRevB.83.134423} although that would entail
designing a device with equal capacitive coupling between all QD
pairs. Importantly, the only interaction among the dots is
capacitative and there is no particle exchange from one dot to the
others, i.e., the interdot tunneling is not allowed since this would
destroy the flavor conservation rule.

The recent progress in fabricating highly tunable TQDs
\cite{Amaha20081322,PhysRevB.77.193306,PhysRevB.79.085124,
PhysRevLett.97.036807,PhysRevB.82.075304} aims to provide a platform
for testing a variety of predicted novel quantum information
processing functionalities and many-body effects.
\cite{PhysRevLett.86.5188,PhysRevLett.90.166803} The goal of this work
is to analyse the transport and thermodynamic properties of the 
$\SU(3)$ Kondo effect in highly symmetric capacitively-coupled TQD. 
Similar study has very recently been performed in
Ref.~\onlinecite{moca2012}; where comparison can be made, our results
agree with theirs. We also study the effect of local perturbations on
the $\SU(3)$ singlet Kondo state such as asymmetrical lead-dot
couplings, finite interdot tunneling rates, non-equal charging
energies, etc. Generally, these perturbations destroy the $\SU(3)$
singlet Kondo state, however, we propose a way to restore the Kondo
resonance by properly gating the dot levels.
  
In order to investigate in a general framework the different regimes
encountered for the the TQD system we consider the operators of the
$\SU(3)$ Lie algebra which describe the orbital (flavor) degree of
freedom of the electrons. The thermodynamics analysis is performed by
calculating the impurity orbital (flavor) susceptibility
$\chi_\mathrm{imp}(T)$, and the impurity entropy $S_\mathrm{imp}(T)$.
Our results indicate that $\SU(3)$ Kondo physics occurs when the QDs
are tuned to single occupancy, $n=1$, or double occupancy, $n=2$,
which is achieved \textit{away from the particle-hole symmetric
point}, contrary to what happens in the more familiar $\SU(2$) and
$\SU(4)$ Kondo cases where at half filling the Kondo effect is
present. Consequently, the Kondo peak itself is not symmetric with
respect to the chemical potential as visible in the spectral densities
for each dot.  These values of the occupancy have important
consequences for the linear conductance. In accordance with the
Friedel-Langreth sum rule, the linear conductance is $G=G_0
\sin^2\delta$ where the scattering phase shift $\delta$ is
approximately given by $\delta =\pi n/N$; here $n=\expv{\hat{n}}$ is
the total TQD occupation and $N=3$, while $G_0$ is defined as
$G_0=e^2/h$ (note that we are considering a spinless system, thus the
spin factor 2 is not present in $G_0$). Therefore, in the $\SU(3)$
Kondo regime with $n=1$ and $n=2$ one has
\cite{PhysRevLett.106.106401}
\begin{equation}
G_\mathrm{Kondo} = \frac{3}{4} e^2/h\,.
\end{equation}
This result must be compared to that at the p-h symmetric point where 
$\delta=\pi/2$, and thus \cite{zinnjustin1998, parcollet1998}
\begin{equation}
G_{\mathrm{p-h}}=e^2/h\,.
\end{equation}

This paper is organized as follows. In Sec. I we introduce the model
Hamiltonian to describe the TQD setup and discuss the theoretical
tools to solve it.  Section II is devoted to the study of the
emergence of the $\SU(3)$ Kondo regime as a function of various
parameters, namely the dot level position ($\epsilon_i$), the interdot
Coulomb interaction ($V_i$), and the lead-dot tunneling couplings
($\Gamma_i$).  The discussion is based on the thermodynamics and we
investigate the behavior of the impurity orbital susceptibility
$\chi_\mathrm{imp}(T)$ and the impurity entropy $S_\mathrm{imp}(T)$.
In Sec. III we list some signatures of the $\SU(3)$ Kondo state in the
transport measurements. In Sec. IV we study the robustness of the
$\SU(3)$ orbital Kondo singlet against diverse perturbations, namely,
asymmetric lead-dot tunneling couplings, different on-site energy
values and possible leaking effects described by nonzero inter-dot
tunnelling rates. In Sec. V we briefly consider the generalization to
the spinfull problem and discuss the different kinds of the Kondo
effect expected in that case. Finally our main conclusions are
summarized in Sec. VI.

\section{Model and methods}

We model the TQD system (see Fig.~1) using a Hamiltonian consisting of
three copies of the non-interacting resonant-level model (each
describing one QD and the effective single channel of the electrons
that the dot hybridizes with) and a coupling term which includes the
inter-dot interactions and any possible inter-dot tunneling:
\begin{equation}
\label{H}
H = \sum_{i=1}^3 H_i + H_\mathrm{int},
\end{equation}
with
\begin{equation}
H_i = \sum_k \epsilon_k c^\dag_{k,i} c_{k,i} + \epsilon_i d^\dag_i d_i
+ v_i \sum_k \left( c^\dag_{k,i} d_i + \text{h.c.} \right),
\end{equation}
and
\begin{equation}
H_\mathrm{int} = 
\sum_{\langle i,j \rangle}
\left[ 
V_{ij} n_i n_j +
t_{ij} \left( d^\dag_i d_j + \text{h.c.} \right) 
\right].
\end{equation}
Here $c^\dag_{k,i}$ is the creation operator for an electron with
momentum $k$ in channel $i$, while $d^\dag_i$ is the creation operator
for an electron in dot $i$; the occupancy operator is defined as $n_i
= d_i^\dag d_i$. Assuming flat conduction bands, the hybridisation of
each channel to the attached dot is characterized by a single number,
$\Gamma_i=\pi \rho v_{i}^2$, where $\rho$ is the density of states in
the band which we take to be constant and of width $2D$ (flat-band
approximation). 
Hereafter, we consider all energies in units of the half-bandwidth,
$D=1$.
$V_{ij}$ is the charge repulsion between two dots,
while $t_{ij}$ is the hopping amplitude between two dots. For
symmetrical configurations we simplify the notation as $\epsilon
\equiv \epsilon_i$, $t \equiv t_{ij}$, $V \equiv V_{ij}$. Notice that
this model is similar to the Coqblin-Schrieffer $\SU(N)$ model. The
Coqblin-Schrieffer model\cite{coqblin1969} describes an impurity in
the $N=2j+1$ representation of the $\SU(N)$ total angular momentum
group. The equivalence is established on low temperature scales where
the charge fluctuations are quenched. Assuming that the conduction
bands are particle-hole (p-h) symmetric ($\epsilon_{-k} =
-\epsilon_k$), the p-h transformation ($d_i^\dag \to d_i$, $d_i \to d_i^\dag$,
$c^\dag_{k} \to c_{-k}$, $c_k \to c^\dag_{-k}$, etc.) leads to (up to
irrelevant constants)
\begin{equation}
{\tilde H}_i = \sum_k \epsilon_k c^\dag_{k,i} c_{k,i}
+ (-\epsilon_i) d^\dag_i d_i + 
(-v_i) \sum_k \left( c^\dag_{k,i} d_i + \text{h.c.} \right)\,,
\end{equation}
and
\begin{equation}
{\tilde H}_\mathrm{int} = \sum_{\expv{i,j}} \left[
V_{ij} (1-n_i) (1-n_j) - t_{ij} \left( d^\dag_i d_j + \text{h.c.} \right)
\right].
\end{equation}
Therefore 
\begin{equation}
{\tilde\epsilon}_i = -\epsilon - \sum_{\expv{i,j}} V_{ij}\,.
\end{equation}
The model is p-h symmetric only for $t_{ij}=0$, since finite inter-dot
hopping breaks the bipartiteness. (The sign change of the
hybridization $v_i$ is of no physical consequence.) For a symmetric
configuration, the model is p-h symmetric around the point
$\epsilon=-V$. Therefore the parameter $\delta=\epsilon+V$ is a
measure of the departure from the p-h symmetry.
It must be emphasized that the $\SU(3)$ Kondo effects that are
discussed in the following occur away from the p-h symmetric point
where the impurity charge fluctuates.  This implies that the Kondo peak
itself is not symmetric (confirmed by numerics, see below).

The standard (Gell-Mann) parametrisation for the generators of
the $\SU(3$) Lie algebra is $K_a=\lambda_a/2$ with $\lambda_a$
being Gell-Mann matrices:
\begin{equation}
\label{eqGM}
\begin{split}
\lambda_1 &= 
\begin{pmatrix}
0 & 1 & 0 \\
1 & 0 & 0 \\
0 & 0 & 0 
\end{pmatrix}
\quad
\lambda_2 =
\begin{pmatrix}
0 & -i & 0 \\
i & 0 & 0 \\
0 & 0 & 0 
\end{pmatrix}
\quad
\lambda_3 = 
\begin{pmatrix}
1 & 0 & 0 \\
0 & -1 & 0 \\
0 & 0 & 0 
\end{pmatrix}
\\
\lambda_4 &=
\begin{pmatrix}
0 & 0 & 1 \\
0 & 0 & 0 \\
1 & 0 & 0 
\end{pmatrix}
\quad
\lambda_5 =
\begin{pmatrix}
0 & 0 & -i \\
0 & 0 & 0 \\
i & 0 & 0 
\end{pmatrix}
\quad
\lambda_6 = 
\begin{pmatrix}
0 & 0 & 0 \\
0 & 0 & 1 \\
0 & 1 & 0 
\end{pmatrix}
\\
\lambda_7 &=
\begin{pmatrix}
0 & 0 & 0 \\
0 & 0 & -i \\
0 & i & 0 
\end{pmatrix}
\quad
\lambda_8 =
\frac{1}{\sqrt{3}}
\begin{pmatrix}
1 & 0 & 0 \\
0 & 1 & 0 \\
0 & 0 & 2 
\end{pmatrix}
\end{split}
\end{equation}
We thus define the $\SU(3)$ operators for the TQD system as
\begin{equation}
\begin{split}
O_a^{(k)} &= \sum_{ij} c^\dag_{k,i} \left( K_a \right)_{ij}
c_{k,j}, \\
O_a^\mathrm{imp} &= \sum_{ij}  d^\dag_i \left( K_a \right)_{ij} d_j, \\
O_a^\mathrm{total} &= O_a^\mathrm{imp} + \sum_k O_a^{(k)},
\end{split}
\end{equation}
where $a=1,\ldots,8$, while $i$ and $j$ range over the three channels,
and $k$ ranges over all conduction-band momenta.
The Casimir operator of $\SU(3)$ is defined as
\begin{equation}
K^2_\mathrm{total} = \sum_{a=1}^{8} (O^\mathrm{total}_a)^2.
\end{equation}
In a fully $\SU(3)$ symmetric case, the traces $\Tr(O_a^2)$ are all
equivalent. In numerical calculations, it is thus sufficient to
calculate the expectation value of a single $O_a^2$ operator; the most
convenient choice is $O_3^2$.
The expectation value of $\expv{K^2_\mathrm{total}}$ is then 8 times this value.

The behavior of an impurity system can be analyzed by studying its
thermodynamic properties. In the following section we will consider
the impurity orbital susceptibility $\chi_\mathrm{imp}(T)$ and the
impurity entropy $S_\mathrm{imp}(T)$. These two quantities 
serve to establish the range of parameters for which the $\SU(3)$ spin 
Kondo physics is encountered.
In the fundamental representation of $\SU(3)$ one has $\expv{K^2}=4/3$.
In the high-temperature regime where all eight dot states are equally
probable, one has $\expv{K^2}=(6 \times 4/3 + 2 \times 0)/8=1$, since
there are six singly occupied states (by either one electron or by one
hole) and two states corresponding to totally empty and totally full
system. The impurity $\SU(3)$ orbital susceptibility (more precisely,
this is the impurity contribution to the total system orbital
susceptibility) is defined as
\begin{equation}
\chi_\mathrm{imp}(T)=\beta \left( \expv{K^2_\mathrm{total}}(T) -
\expv{K^2_\mathrm{total}}_0(T) \right),
\end{equation}
where the bracket with subscript 0 denotes the result for the system
without the dots (i.e., the Hamiltonian $H$ consists only of the
conduction bands). Here $\beta=1/k_B T$ with $k_B$ the Boltzmann
constant. The value of $k_B T \chi(T)$ therefore indicates the
presence of a finite effective orbital \textit{local moment} on the
TQD and it can be used to classify the fixed points
\cite{krishna1980a, krishna1980b}.

The impurity entropy is a measure of the number of the effective
degrees of freedom of the TQD at a given parameter configuration.
It is defined through 
\begin{equation}
S_\mathrm{imp}(T) = \frac{(E-F)}{T} - \frac{(E-F)_0}{T},
\end{equation}
where $E=\expv{H}=\Tr[H \exp(-H/k_B T)]$ and $F=-k_B T \ln
\Tr[\exp(-H/k_B T)]$.

We also compute the dot spectral functions $A(\omega,T)$ and compute
the differential conductance through each dot using the Meir-Wingreen formula as
\cite{meir1992}
\begin{equation}
G(T)=G_0 \int_{-\infty}^{\infty} \left( -\frac{\partial f}{\partial
\omega} \right)
\pi \Gamma A(\omega,T) \mathrm{d}\omega,
\end{equation}
where $G_0=e^2/h$ and $f=[1+\exp(\omega/k_B T)]^{-1}$ is the
Fermi-Dirac distribution function; the chemical potential has been
fixed at zero energy.

The calculations have been performed using the numerical
renormalization group method
\cite{wilson1975,krishna1980a,krishna1980b,bulla2008} as implemented
in the ``NRG Ljubljana'' code. We have used the discretization
parameter $\Lambda=8$ with the $z$-averaging over $N_z=8$ values. We
have verified that such a large value of $\Lambda$ still
produces reliable results by performing a convergence study as a
function of $\Lambda$ down to $\Lambda=2$. In the NRG truncation, we
have kept states with energy up to $10\omega_N$ where $\omega_N$ is
the characteristic energy scale at the $N$-th NRG step, or at most
6000 states. For calculating the spectral functions, we have used the
complete Fock space method \cite{peters2006,weichselbaum2007}. Very
recently, a study of the fully symmetric SU(3) model has been
performed with an implementation of the NRG which can explicitly use
the SU(3) symmetry of the model to simplify the calculations
\cite{moca2012}. Here we only use the U(1) total-charge-conservation
symmetry, thus the calculations are significantly more time-demanding.
However, our approach makes it possible to study the effects of the
symmetry breaking terms, which is important for physical realizations
of this model.

\section{Numerical results: valence fluctuating and the $\SU(3)$ Kondo regimes}

\begin{figure}[htbp]
\centering 
\includegraphics[clip,width=8cm]{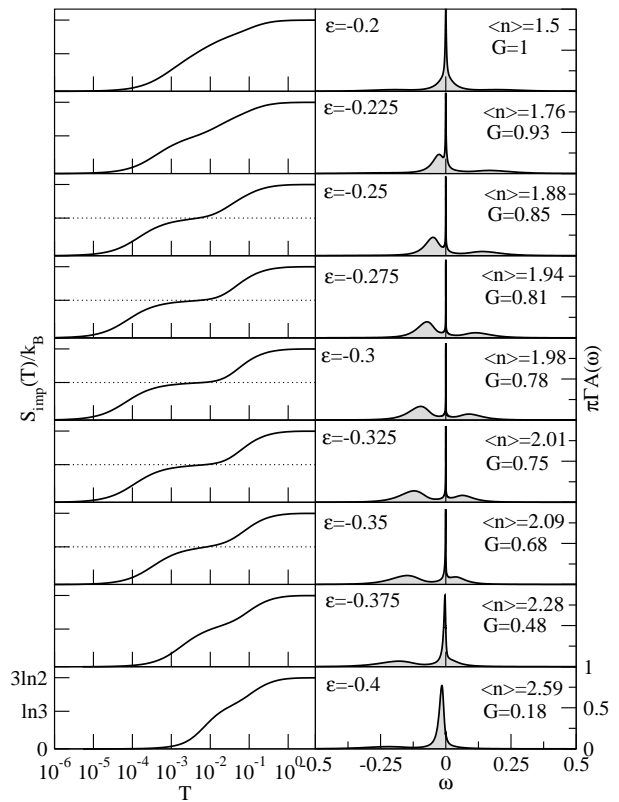}
\caption{Temperature dependence of the impurity entropy (left panels)
and the zero-temperature dot spectral function (right panels) for a range of the
on-site energies $\epsilon$. We consider a symmetric triple quantum dot system. 
The total occupancy of the triple quantum dot, $\langle
n \rangle$, and the zero-temperature linear conductance through one dot, $G$,
are also shown. The interdot tunnelling is zero, $t=0$. Other
parameters are $\Gamma=0.01$, $V=0.2$.}
\label{fig1}
\end{figure}

\begin{figure}[htbp]
\centering 
\includegraphics[clip,width=8cm]{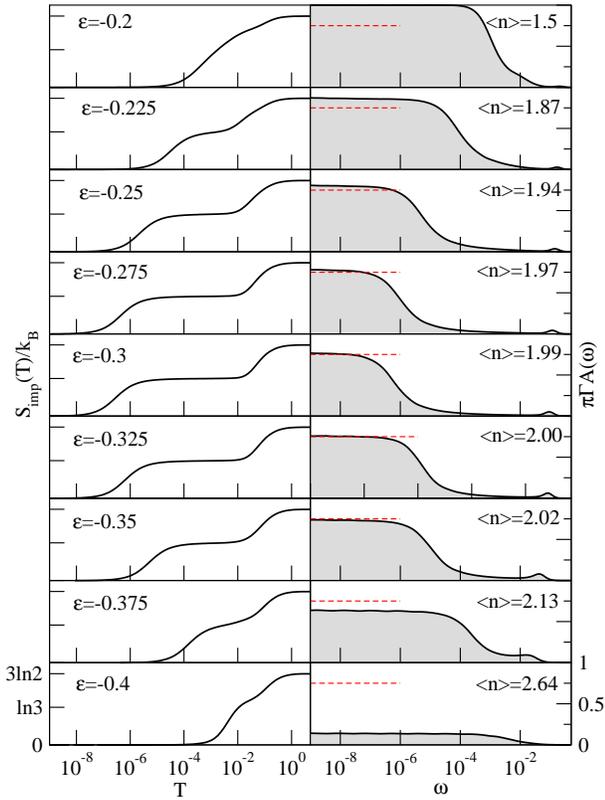}
\caption{(Color online) Temperature dependence of the impurity entropy (left panels)
and the zero-temperature dot spectral function (right panels) for a range of the
on-site energies $\epsilon$. Same parameters as in Fig.~\ref{fig1},
but with smaller hybridization, $\Gamma=0.005$. We plot the
positive-frequency side of the spectral function on the logarithmic
frequency scale. The dashed line corresponds to the characteristic
conduction of $G=(3/4)(e^2/h)$.}
\label{fig1bis}
\end{figure}

In order to identify the different regimes of the TQD system we 
consider the thermodynamic and transport properties. In
Fig.~\ref{fig1} we show the basic results for the fully $\SU(3)$
symmetric case where all the dots and hybridizations are equivalent.
The lead-dot couplings have common value of $\Gamma=0.01$, and there
is no inter-channel tunneling (i.e., $t \equiv 0$).  Since the system
is symmetric with respect to the point $\epsilon=-V$ for this choice
of parameters, we consider only the value of the on-site energy
$\epsilon$ below $-V$ (i.e., $\delta<0$); other results can be
obtained by an appropriate p-h transformation.

We observe that for a range of on-site energies $-0.35 \lesssim
\epsilon \lesssim -0.25$, the occupancy reaches values close to $2$.
In this case, the system evolves from the \textit{free-orbital} fixed
point (fp) with impurity entropy of $3\ln 2$ to a \textit{local-moment} fp
with triple degeneracy (indicated by a $\ln3$ plateau in the
impurity entropy) as the temperature decreases below the
charge-fluctuation scale of $~V$. The triple degeneracy is finally
lifted at low temperatures and then we reach a \textit{non-degenerate
strong-coupling} fp corresponding to the $\SU(3)$ Kondo
regime. In this regime the zero-temperature linear conductance is near
$3/4$, as predicted in Ref.~\onlinecite{PhysRevLett.106.106401} based on the
Friedel sum rule arguments [see Eq. (1)].

A very different behavior is found near the p-h symmetric point at
$\epsilon=-V$. Here the system evolves from the \textit{free-orbital}
to the \textit{valence-fluctuation} fp with entropy $\ln 6$
(only visible as a weak bulge in the $S_\mathrm{imp}(T)$ curve in
Fig.~\ref{fig1}).  In this case the valence-fluctuation regime
corresponds to charge fluctuations from $n=1$ to $n=2$ charge states.
The entropy is eventually reduced from $\ln 6$ to zero at some low
temperature. In this case the entropy is released as the system
evolves to the \textit{strong-coupling} fp without passing
through the local-moment fp.

By reducing the dot-lead hybridisation by half, i.e., for
$\Gamma=0.005$, as shown in Fig.~\ref{fig1bis}, the $\SU(3)$ Kondo
regime is even more clearly discernible and we can see that the occupancy
is pinned to the value 2 for a much broader range of dot potential
energies.  Here, the conductance reaches the universal value of $G=3/4
(2e^2/h)$ for a wide range of $\epsilon$ due to a much more
robust $\SU(3)$ Kondo state.

\begin{figure}[htbp]
\centering 
\includegraphics[clip,width=8cm]{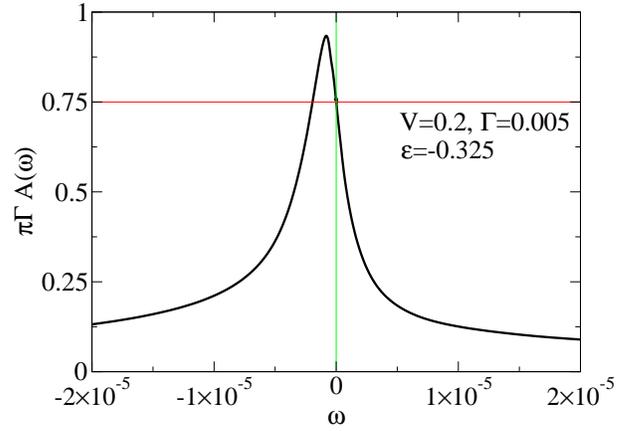}
\caption{(Color online) Spectral function in the $\SU(3)$ Kondo regime. The triple
quantum dot system is symmetric and the total occupancy is $\langle n
\rangle$=2. Note the pinning of the zero-frequency spectral density to
the value $A(0)=(3/4)/\pi\Gamma$ and the important fact that the actual Kondo peak is
displaced away from $\omega=0$.} \label{fig9}
\end{figure}

In the $\SU(3)$ Kondo regime, the dot spectral density,
shown in Fig.~\ref{fig9}, displays a Kondo resonance with a maximum
height shifted away from the Fermi level. As
previously noticed this is the result of having the $\SU(3)$ Kondo
regime away from the p-h symmetric point. In addition, the
shifted spectral density produces a zero-temperature linear
conductance $G=3/4 (e^2/h)$, which is by itself a hallmark of the
occurrence of the $\SU(3)$ Kondo physics. This is in contrast with the
$\SU(4)$ case where the linear conductance coincides in value with
the linear conductance for the $\SU(2)$ Kondo effect.

\begin{figure}[htbp]
\centering 
\includegraphics[clip,width=8cm]{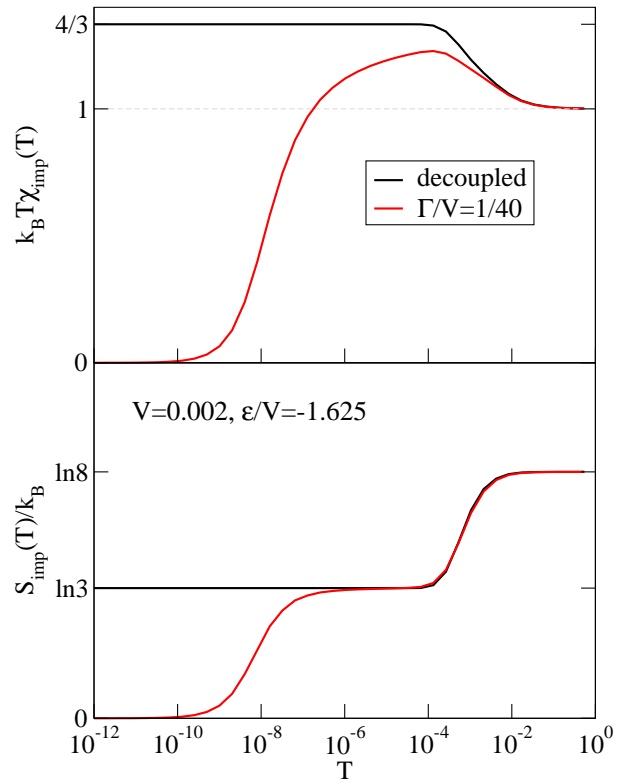}
\caption{(Color online) Thermodynamic properties in the $\SU(3)$ Kondo regime. 
We consider a symmetric TQD system.
The total occupancy of the triple quantum dot, $\langle
n \rangle=2$. We compare the cases of a decoupled triple 
quantum dot system ($\Gamma=0$) and the triple quantum dot 
connected to the leads ($\Gamma=0.005$) in which the Kondo correlations
are present.}
\label{fig10}
\end{figure}

The thermodynamic properties of the $\SU(3)$ Kondo regime are studied
in more detail in Fig.~\ref{fig10}. In order to gain some intuition
about the role of the Kondo correlations in the $\SU(3)$ orbital
susceptibility we compare the case of an uncoupled TQD system, where
all three lead-dot hybridizations are $\Gamma=0$, with the case where
the TQD is connected to leads and the $\SU(3)$ Kondo state builds up.
In the high-temperature limit, in both cases the spin susceptibility
is $1$ and all 8 TQD states are equally probable, thus there is $\ln
8=3\ln 2$ impurity entropy.  As the temperature decreases below $V$,
so that the charge fluctuations are frozen out, the $\SU(3)$
susceptibility of $4/3$ is established, as expected for the
fundamental triplet representation of $\SU(3)$. This is the
\textit{local-moment} fp.  The decoupled system stays in this fixed
point down to $T=0$, while at finite $\Gamma$ the local moment is
screened in the $\SU(3)$ Kondo effect and the susceptibility vanishes,
as expected.  At this point the system is in the
\textit{non-degenerate strong-coupling} fp in which the ground state
corresponds to a $\SU(3)$ Kondo singlet state. Notice that the
transition from the \textit{free-orbital} to the \textit{local-moment}
regime and then eventually to the \textit{strong-coupling regime} is
fully analogous to the behaviour in the standard single-impurity
Anderson model with the $\SU(2)$ symmetry
\cite{wilson1975,krishna1980a,krishna1980b}. We also emphasize that
the low-temperature parts of the impurity susceptibility and impurity
entropy are universal and that the scaling of the results for
different parameters is observed if the temperature axis is rescaled
by an appropriately defined Kondo temperature $T_K$ (see below).

\begin{figure}[htbp]
\centering
\includegraphics[clip,width=8cm]{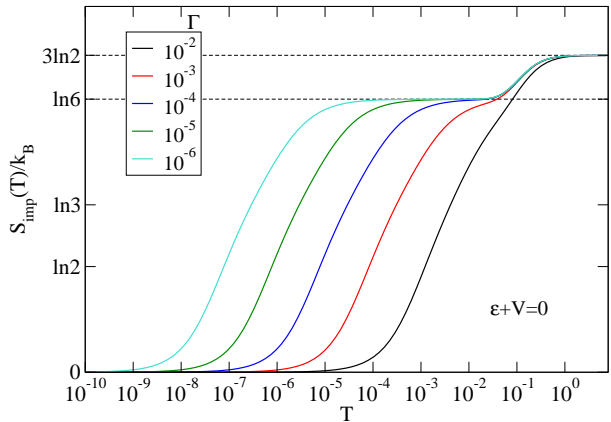}
\caption{(Color online) Temperature dependence of the entropy at the
particle-hole symmetric point for a range of hybridization parameters
$\Gamma$. We consider a symmetric triple quantum dot system with
parameters $V=0.2$, $\epsilon=-0.3$. } \label{fig6}
\end{figure}

For completeness we also analyse the p-h symmetric point, in which the
only low-temperature scale is $\Gamma$ itself and there is no
Kondo-like screening. In this model, the p-h symmetric point
corresponds to a valence-fluctuation regime where charge fluctuations
occur. In Fig.~\ref{fig6} the temperature dependence of the 
entropy is shown for a symmetric TQD and various $\Gamma$ values when
$\epsilon=-V$. At high temperatures the TQD is found in the
free-orbital regime where the TQD entropy is $\ln 8$.  Then, the
system crosses over on the temperature scale of $V$ to a
valence-fluctuation fp with a six-fold degenerate ground state in
which the entropy reaches the value of $\ln 6$. In this case, there
can be either a single electron or a single hole in the three dots for
a total of six states with the same energy. Decreasing further the
temperature the system crosses over to the non-degenerate ground state
with zero entropy at the temperature scale of $\Gamma$, see
Fig.~\ref{fig6}. There is no further dynamically generated low-energy
scale in this case.

\begin{figure}[htbp]
\centering
\includegraphics[clip,width=8cm]{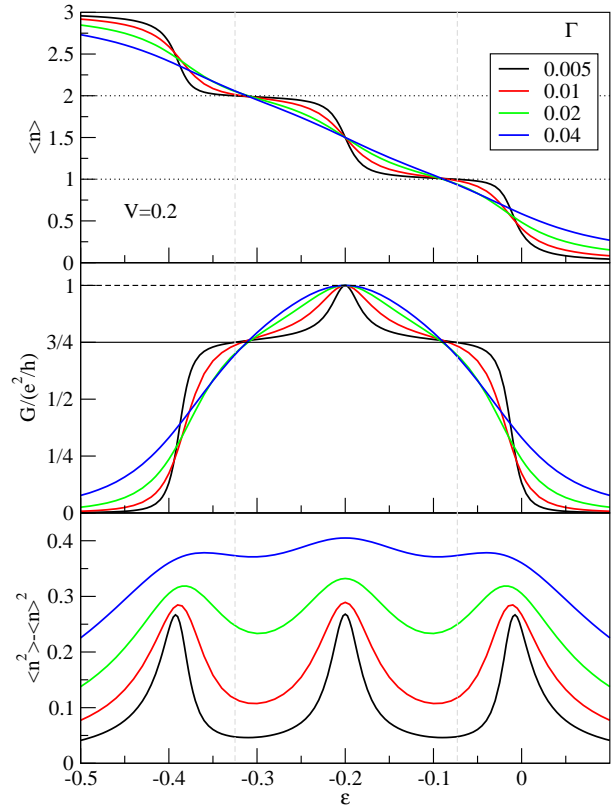}
\caption{(Color online) Occupancy $n$, differential conductance $G$
and charge fluctuations $\delta n^2 = \langle n^2 \rangle - \langle n
\rangle^2$ as a function of the on-site energy for a range of
hybridizations $\Gamma$.} \label{newnrg1}
\end{figure}

The zero-temperature fixed points for different $\epsilon$ form a line
of fixed points which are related by the different strength of the
residual potential scattering experienced by the quasi-particles. For
an overview of the system behavior, in Fig.~\ref{newnrg1} we plot the
zero-temperature total TQD occupancy, linear conductance, and charge
fluctuations as a function of the on-site energy $\epsilon$ for
several choices of the hybridization $\Gamma$. The emergence of the
Kondo plateau for low enough $\Gamma$ is clearly visible; it coincides
with the regions of low charge fluctuations in the TQD.

\section{Evidence of $\SU(3)$ Kondo correlations in transport measurements}

% TO DO? plot Re Sigma and Im Sigma
% TO DO? section on EOMs for this system

\begin{figure}[htbp]
\centering
\includegraphics[clip,width=8cm]{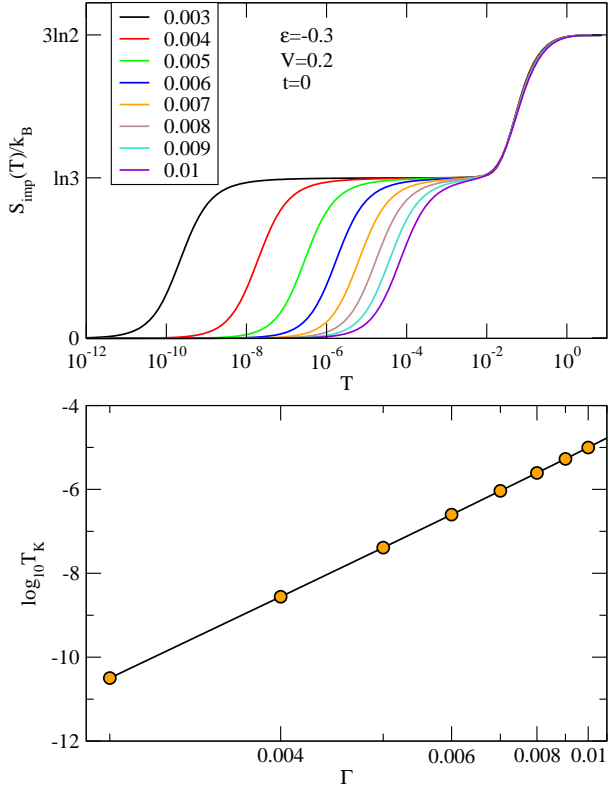}
\caption{(Color online) Top panel: Temperature dependence of the
entropy for a symmetric triple quantum dot system for a range of
hybridisation parameters $\Gamma$. Other parameters are tuned so that
the system is in the $\SU(3)$ Kondo regime, $V=0.2$, $\epsilon=-0.3$.
Bottom panel: Relation between the Kondo temperature $T_K$ and the
hybridisation $\Gamma$. Here we use an arbitrary definition of the
Kondo temperature, $S_\mathrm{imp}(T_K)=0.1k_B$.} \label{fig2}
\end{figure}

In the previous section, we have demonstrated 
the occurrence of the $\SU(3)$ Kondo effect by considering
the thermodynamic properties. Usually, one way to probe 
the existence of Kondo correlations in QD
systems is to measure the exponential dependence of the Kondo energy
scale ($k_B T_K$) 
 with the inverse hybridization $1/\Gamma$ as
\begin{equation}\label{kondotem}
\ln T_K \propto -\frac{1}{\Gamma}.
\end{equation}
In the NRG calculations this dependence is demonstrated by using a
$T_K$ defined from the entropy curve as $S_\mathrm{imp}(T_K)=0.1k_B$,
see the upper panel of Fig.~\ref{fig2}. Plotting $T_K$ for different $\Gamma$
we uncover the exponential dependence, see the lower panel of
Fig.~\ref{fig2}.

In view of this result, the occurrence of $\SU(3)$ Kondo correlations
in a TQD experiment could be demonstrated by performing transport
measurements, for instance by measuring the Kondo temperature $T_K$.
Usually this is achieved by identifying the half width at the half
maximum (HWHM) of the non-linear conductance peak through one of the
dots with $T_K$. This measurement would, however, be rendered
problematic in this system due to the asymmetric shape of the Kondo
resonance. In $\mathrm{d}I/\mathrm{d}V$ measurement with finite bias,
the current is namely given approximately by
\begin{equation}
I(T,V) = \int [f_L(\omega)-f_R(\omega)] \pi \Gamma A(\omega,T)
\mathrm{d}\omega,
\end{equation}
where we have neglected the voltage-dependence of the spectral
function, and we assume $f_L(\omega)=f(\omega-V/2)$ and
$f_R=f(\omega+V/2)$. In the zero-temperature limit, we thus find
approximately
\begin{equation}
\frac{\mathrm{d}I}{\mathrm{d}V} = G_0 \pi \Gamma 
\frac{A(V/2,0) + A(-V/2,0)}{2}.
\end{equation}
Since the spectral function $A(\omega)$ is asymmetric, care is needed
in extracting the width of the spectral function peak from the
differential conductance measurements at finite bias voltage (even
when the non-equilibrium effects are neglected). When properly done, by
measuring $T_K$ and then varying $\Gamma$ one should be able to get a
relation that matches Eq.~\eqref{kondotem}. 

\begin{figure}[htbp]
\centering
\includegraphics[clip,width=8cm]{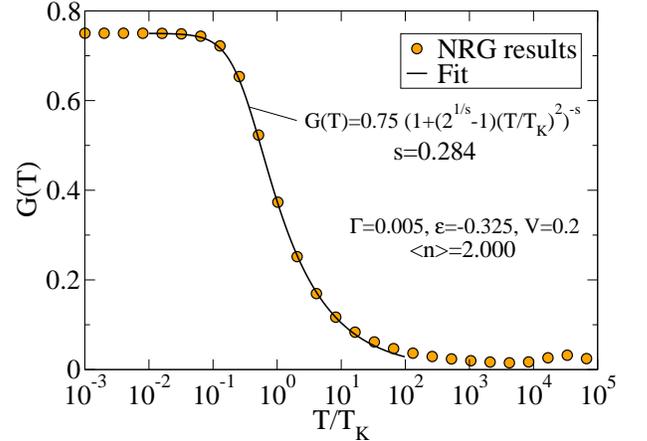}
\caption{(Color online) Temperature dependence of the linear
conductance in the $\SU(3)$ Kondo regime of a symmetric
triple quantum dot.} \label{fig8}
\end{figure}

Another signature of the $\SU(3)$ Kondo physics could be detected
through the temperature dependence of the linear conductance $G$.
Close to zero temperature, $G$ is $3/4$ in units of the conductance
quantum $G_0=e^2/h$ (note the absence of factor 2 in this spinless
case).  At small but finite temperatures, we find that the
low-temperature conductance fits, similarly to the $\SU(2)$ Kondo case,
to the empirical formula of the general form
\cite{goldhabergordon1998a}
\begin{equation}
\label{eqG}
G(T)=G_0 (1+(2^{1/s}-1)(T/T_K)^2)^{-s},
\end{equation}
with $s=0.23$ for the $\SU(2)$ case and with $s=0.28$ for the $\SU(3)$ Kondo case. The value of $s=0.28$
has been extracted from the NRG results for $G(T)$. The fitting to $G(T)$ is performed for 
 a symmetric TQD at $\epsilon=-0.2$, and $\Gamma=0.005$, see
Fig.~\ref{fig8}. It is interesting to notice that a single fit
formula, Eq.~\eqref{eqG}, applies over many orders of magnitude in
temperature in a number of quantum impurity models that exhibit Kondo
effects of very different kinds; see, for example,
Refs.~\onlinecite{goldhabergordon1998a,parks2010}.

\section{Departure from the $\SU(3)$ strong-coupling fixed point}

In real experiments it is extremely challenging to construct perfectly
symmetric multi-dot systems. Therefore, in order to experimentally
detect $\SU(3)$ Kondo correlations we need to know to what extent the
$\SU(3)$ Kondo physics is robust against all possible local
perturbations. These perturbation can arise from asymmetric lead-dot
tunneling couplings, possible tunneling events among the dots, and
different on-site dot level potentials or distinct inter-dot Coulomb
energies.  In spite of the presence of unavoidable local perturbations
in real setups that eventually destroy the $\SU(3)$ Kondo state, below
we give a protocol to restore the Kondo correlations by properly
adjusting the dot potentials.

\subsection{Asymmetric lead-dot hybridizations}

\begin{figure}[htbp]
\centering
\includegraphics[clip,width=8cm]{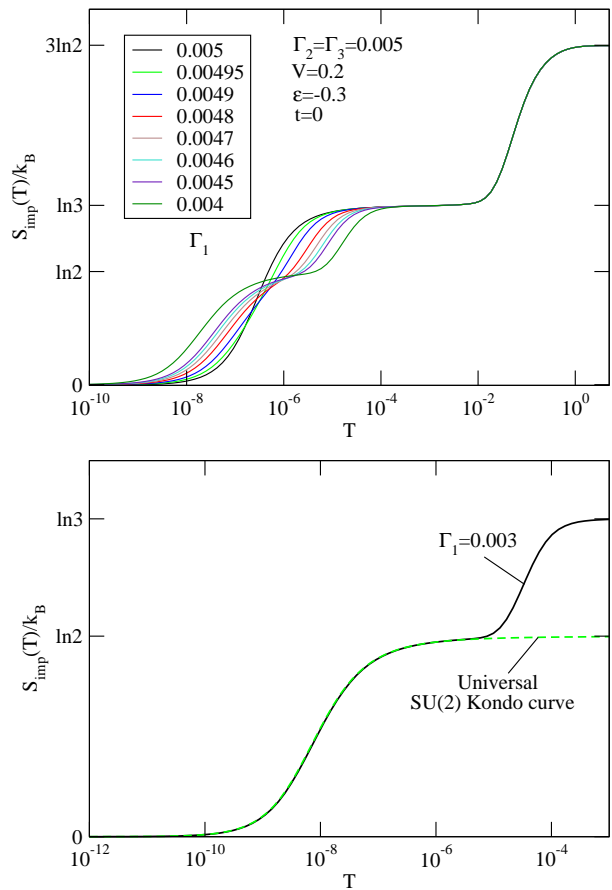}
\caption{(Color online) Top panel: Temperature dependence of the
impurity entropy for a range of the hybridization parameter
$\Gamma_1$, while $\Gamma_2=\Gamma_3=\Gamma$ is held fixed. Bottom
panel: Impurity entropy for a strong symmetry breaking and a fit of
the $\SU(2)$ Kondo screening cross-over with the universal $\SU(2)$
entropy curve for the spinfull single dot case.}
\label{fig3}
\end{figure}

First, we analyse the effect of having asymmetric lead-dot coupling.
If one of the hybridizations $\Gamma_i$, say $\Gamma_1$, is made
weaker, the $\SU(3)$ symmetry is broken. Even small changes of
$\Gamma_1$ are sufficient; the effect is similar to the induced
magnetization by a ferromagnetic conduction band in the spinfull
single QD device. In the upper panel of Fig.~\ref{fig3} we show the
impurity entropy evolution as the temperature is lowered for constant
$\Gamma_2=\Gamma_3=\Gamma$ and $\Gamma_1$ ranging from
$\Gamma_1=\Gamma$ (symmetric configuration) to $\Gamma_1=0.9\Gamma$
(asymmetric lead-dot couplings).  When the temperature is lowered, in
the asymmetric lead-dot configuration the system flows from the
$\SU(3)$ local-moment fp (with impurity entropy $\ln 3$) to a new
$\SU(2)$ local-moment fp with two-fold degeneracy (with impurity
entropy $\ln 2$). The $\SU(2)$ local moment is then screened in the
conventional $\SU(2)$ Kondo effect which lifts the degeneracy at
temperatures well below the new $\SU(2)$ Kondo energy scale (see lower
panel in Fig.~\ref{fig3} for the comparison between the entropy
evolution for the asymmetric coupled triple dot case and the universal
$\SU(2)$ Kondo model). If, however, $\Gamma_1$ is increased above
$\Gamma_2=\Gamma_3$ rather than decreased, the system crosses over
from the $\SU(3)$ local-moment fp to the frozen-impurity fp without
any Kondo screening (results not shown). The same happens when all
three $\Gamma_i$ are different.

\begin{figure}[htbp]
\centering
\includegraphics[clip,width=8cm]{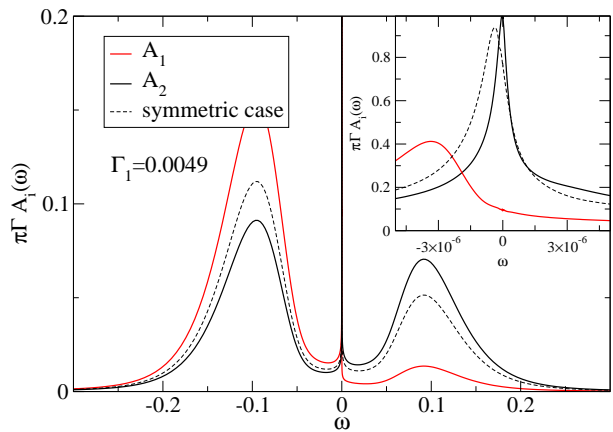}
\caption{(Color online) Spectral functions in the case of non equal hybridisations
$\Gamma_i$. The dashed line corresponds to the fully symmetric case
with $\Gamma \equiv \Gamma_i = 0.005$.}
\label{fig3bis}
\end{figure}

As already mentioned, the asymmetry of the lead-dot couplings in the
$\SU(3)$ Kondo effect is analogous to the $\SU(2)$ Kondo physics in
the presence of ferromagnetic contacts. In the latter case, the Kondo
resonance is split due to the appearance of an induced exchange field
because of the polarized contacts. The same physical behavior is
obtained for the $\SU(3)$ Kondo case. Figure~\ref{fig3bis} shows this
result. We plot the spectral densities for the dot 1 and dot 2,
denoted as $A_{1}(\omega)$ and $A_{2}(\omega)$. We consider the case
of symmetric couplings ($\Gamma=\Gamma_1=\Gamma_2=\Gamma_3$) in which
$A_{1}(\omega)=A_{2}(\omega)$ and the case of asymmetrical lead-dot
coupling configuration where $\Gamma_2=\Gamma_3=0.005$ and
$\Gamma_1=0.003$. In the latter case the two spectral functions show
$\SU(3)$ Kondo-peak splitting which is better visible in the close-up
shown as an inset in Fig.~\ref{fig3bis}.  It is noteworthy that the
Kondo spectral peaks for $A_2=A_3$ reach a high value (approaching, in
fact, the unitary limit), while that for $A_1$ is strongly suppressed.
This is related to the fact that the dots 2 and 3 are $\SU(2)$ Kondo
screened, thus their zero-bias conductance remains nearly unitary,
while the dot 1 becomes decoupled and it is only weakly conducting.

\subsection{Unequal on-site dot energies}

\begin{figure}[htbp]
\centering
\includegraphics[clip,width=8cm]{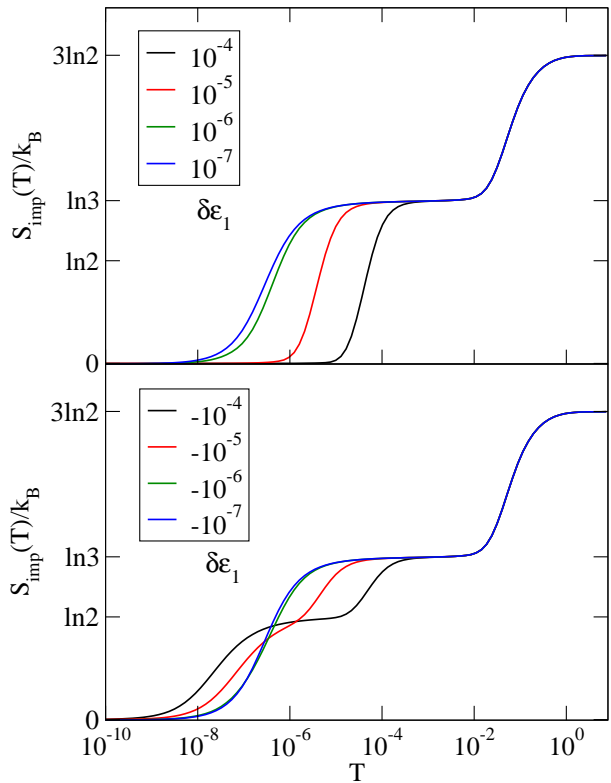}
\caption{(Color online) Temperature dependence of the impurity 
entropy for a range of the parameter $\delta\epsilon_1$ defined as
$\epsilon_1=\epsilon+\delta\epsilon_1$, while $\epsilon_2=\epsilon_3=\epsilon$
with $\epsilon=-0.3$. The rest of the parameters correspond to a
symmetric triple quantum dot configuration with $V=0.2$,
$\Gamma=0.005$.}
\label{fig11}
\end{figure}

The $\SU(3)$ Kondo state can also be destroyed by having non-equal dot
level energies. This case is illustrated in Fig.~\ref{fig11} where it
is shown how the $\SU(3)$ symmetry is broken by changing the on-site
energy $\epsilon_1$ away from the common value $\epsilon$, i.e., 
$\epsilon_1=\epsilon+\delta \epsilon_1$, with $\delta\epsilon_1$ being the detuning.  The
resulting new state depends on the direction of the detuning. For
positive detuning $\delta\epsilon_1>0$, shown in the upper panel in Fig. ~\ref{fig11}, the
$\SU(3)$ Kondo effect is quenched on the energy scale of the detuning.
However, for negative detuning $\delta\epsilon_1<0$, shown in the
lower panel, the system evolves from the
three fold degenerate local moment fp with $S_\mathrm{imp}=\ln 3$ to a
two fold degenerate local moment fp in which $S_\mathrm{imp}=\ln 2$.
In this case the $\SU(2)$ local moment emerges from two states that
originally formed the $\SU(3)$ triplet local moment.

\subsection{Finite interdot tunneling: $t\neq0$}

\begin{figure}[htbp]
\centering
\includegraphics[clip,width=8cm]{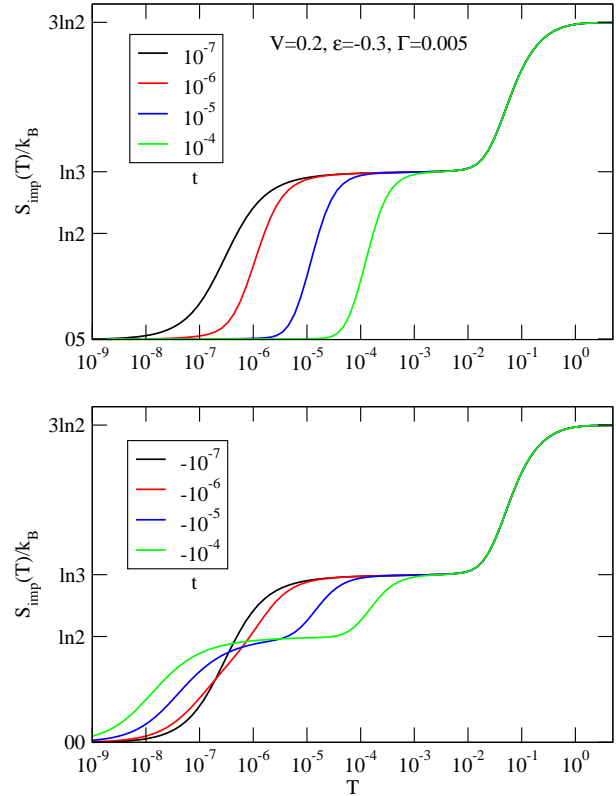}
\caption{(Color online) Temperature dependence of the triple quantum
dot entropy for different values of inter-dot tunneling.}
\label{fig4}
\end{figure}

We have also investigated the fact that the three-fold 
symmetry can also be broken by a finite interdot tunneling. The
behavior yet again depends on the sign of $t$: the ground state may either be a
single state (not Kondo screened) or a two-fold degenerate pair which
is Kondo screened.  The positive $t$ case is considered in the upper
panel of Fig.~\ref{fig4}. Here the entropy evolves from $\ln 3$ to
zero on the temperature scale of $t$ signaling the destruction of the
$\SU(3)$ Kondo singlet.  For negative $t$, the lower panel in the figure
shows that the impurity entropy evolves from $\ln 3$ to
$\ln 2$ on the temperature scale of $|t|$ and the latter corresponds
to local moment fp of two-fold pair of states and it
constitutes a $\SU(2)$ local-moment which undergoes the $\SU(2)$ Kondo
screening at much lower temperatures.

\subsection{Restoration of the SU(3) Kondo physics}

\begin{figure}[htbp]
\centering
\includegraphics[clip,width=8cm]{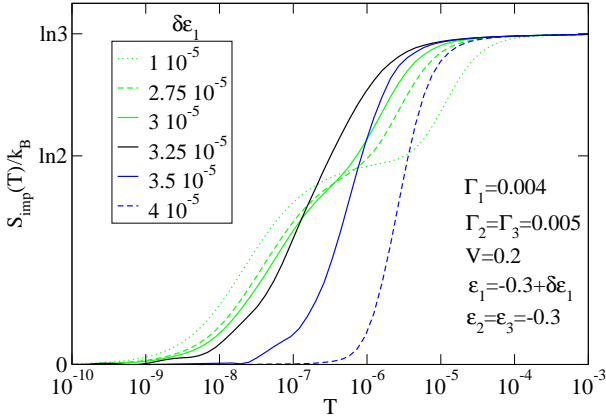}
\caption{(Color online) Restoration of the $\SU(3)$ Kondo effect in an
asymmetrically lead-dot coupled triple quantum dot ($\Gamma_1=0.004$,
and $\Gamma_2=\Gamma_3=0.005$) by tuning the on-site dot potential
$\epsilon_1=\epsilon+\delta\epsilon_1$, with $V=0.2$, and $\epsilon=-0.3$.  The
$\SU(3)$ symmetry is effectively restored for $\delta\epsilon_1=3.25\
10^{-5}$, which is indicated by the impurity curve having the
universal $\SU(3)$ shape. The curves for parameters near the
restoration point show some wiggles; these are numerical artifacts.}
\label{fig5}
\end{figure}

Symmetry can also be broken by having different inter-dot charge
repulsion parameters $V_{ij}$. In experiments,
the capacitive couplings $V_{ij}$ are the most difficult to control,
followed by the inter-dot hopping parameters $t_{ij}$ and hybridizations
$\Gamma_i$, while the on-site energies $\epsilon_i$ are typically the
easiest to tune. There are eight ``directions'' for an $\SU(3)$
symmetry-breaking field, corresponding to the eight generators of this
symmetry. A consideration of the Gell-Mann matrices in
Eq.~\eqref{eqGM} suggest that two ($\lambda_3$ and $\lambda_8$) are
associated with the energy levels, since they are diagonal, while the
remaining six are out-of-diagonal and thus associated with the inter-dot
tunneling. This immediately suggests that symmetry breaking by
tunneling cannot be compensated electrostatically. It is thus
essential, first of all, to ensure that the inter-dot tunneling is
low-enough for the observation of the $\SU(3)$ Kondo effect. In
essence, the tunneling rate should be much smaller than the
anticipated Kondo temperature scale (in appropriate frequency units).
Any remaining asymmetry then corresponds to $\SU(3)$ ``fields'' in
$\lambda_3$ and $\lambda_8$ directions which arise from asymmetric
$\epsilon_i$, $\Gamma_i$ and $V_{ij}$. Assuming that the three on-site
energies $\epsilon_i$ can be freely and independently tuned, it
appears possible to compensate the asymmetries in $\Gamma_i$ and
$V_{ij}$ since there are three parameters to drive two ``fields'' to
zero. We demonstrate this procedure in Fig.~\ref{fig5} where an
asymmetry in the hybridization constants is compensated by tuning the
on-site potentials. Thus, making use of the high degree of tunability
in QD devices, any source of symmetry breaking which naturally arise
from the manufacturing process can be compensated by properly
adjusting the gate voltages.

\section{Spinfull triple quantum dot}

For completeness, in this section we briefly discuss the version of
our model with spin degrees of freedom and additional on-site
electron-electron repulsion terms $U$. The Hamiltonian takes the form
of three copies of the single-impurity Anderson model and a coupling
term with all inter-dot terms:
\begin{equation}
\label{Hspin}
H=\sum_{i=1}^3 H_i + H_\mathrm{int},
\end{equation}
with
\begin{equation}
\begin{split}
H_i &= \sum_{k,\sigma} \epsilon_k c^\dag_{k,\sigma,i} c_{k,\sigma,i}
+ \epsilon_i n_i 
+ U_i n_{\uparrow,i} n_{\downarrow,i}
\\
&+ v_i \sum_{k,\sigma} \left( c^\dag_{k,\sigma,i} d_{\sigma,i} +
\text{h.c.} \right)
\end{split}
\end{equation}
and
\begin{equation}
H_\mathrm{int} = \sum_{\langle i,j \rangle} \left[ V_{ij} n_i n_j + t_{ij}
\sum_\sigma \left( d^\dag_{\sigma,i} d_{\sigma,j} + \text{h.c.}
\right) \right].
\end{equation}
Most terms have the same meaning as in the spinless model,
Eq.~\eqref{H}, but now $n_{\sigma,i} = d^\dag_{\sigma,i} d_{\sigma,i}$
and $n_i = n_{\uparrow,i} + n_{\downarrow,i}$.

Here we will only consider some features of this model. We restrict
our attention to a model with no inter-dot tunneling, $t_{ij} \equiv
0$, and full orbital symmetry, i.e., $\epsilon_i \equiv \epsilon$,
$U_i \equiv U$, $V_{ij} \equiv V$ and $v_i \equiv v$ (or,
equivalently, $\Gamma_i \equiv \Gamma$). Despite the high symmetry of
this problem, it is still too complex for a detailed study using the
NRG technique. For this reason, we resort to a different numerical
approach, the Gunnarson-Sch\"onhammer variational method. Following
Refs.~\onlinecite{Schonhammer76} and \onlinecite{Gunnarson85} we form
a variational ansatz for the ground state wavefunction
$\left|0\right\rangle $ of the Hamiltonian (\ref{Hspin}), 
\begin{eqnarray}
\left|0\right\rangle & = &
\hphantom{+}\sum_{\hphantom{k,}n_{1}n_{2}n_{3}\hphantom{,i}}\lambda_{n_{1}n_{2}n_{3}}P_{n_{1}n_{2}n_{3}}\left|\tilde{0}\right\rangle +\nonumber \\
 & & +\sum_{k,i,n_{1}n_{2}n_{3}}\lambda_{n_{1}n_{2}n_{3}}^{d\rightarrow k,i}P_{n_{1}n_{2}n_{3}}\sum_{\sigma}c_{k,\sigma,i}^{\dagger}d_{\sigma,i}\left|\tilde{0}\right\rangle +\nonumber \\
 & & +\sum_{k,i,n_{1}n_{2}n_{3}}\lambda_{n_{1}n_{2}n_{3}}^{k\rightarrow d,i}P_{n_{1}n_{2}n_{3}}\sum_{\sigma}d_{\sigma,i}^{\dagger}c_{k,\sigma,i}\left|\tilde{0}\right\rangle .\label{eq:gsansatz}
\end{eqnarray}
Here $\left|\tilde{0}\right\rangle $ is the ground state wavefunction
of the noninteracting part ($U=V=0$) of the Hamiltonian (\ref{Hspin}) with
renormalized dot energy levels $\tilde{\epsilon}$ and lead-dot
couplings $\tilde{v}$,
\begin{eqnarray}
\tilde{H}\left(\tilde{\epsilon},\tilde{v}\right)&=&\hphantom{+}\sum_{k,\sigma,i}\epsilon_{k}c_{k,\sigma,i}^{\dagger}c_{k\sigma,i}+\sum_{i}\tilde{\epsilon}n_{i}+\nonumber\\ 
 & & +\sum_{k,\sigma,i}\left(\tilde{v}d_{\sigma,i}^{\dagger}c_{k\sigma,i}+\mathrm{h.c.}\right).\label{flham}
\end{eqnarray}
Projectors $P_{n_{1}n_{2}n_{3}}$ project this state to subspaces with
$n_{i}=0,1,2$ electrons in the $i$-th dot. Variational terms in the
second and the third row of Eq.~(\ref{eq:gsansatz}) provide states
containing an electron above the Fermi energy and a hole below the
Fermi energy in one of the leads, respectively. An approximation to
the true Hamiltonian (\ref{Hspin}) ground state energy
$E_{0}\left(\tilde{\epsilon},\tilde{v}\right)$ and the corresponding
ground state wavefunction coefficients $\lambda$ are calculated by
solving the Schr\" odinger equation within the Hilbert space of the
ansatz. The ground state energy is then further minimized with respect
to $\tilde{\epsilon}\rightarrow\tilde{\epsilon}_{0}$ and
$\tilde{v}\rightarrow\tilde{v}_{0}$, providing us with the
noninteracting part of the Fermi liquid quasiparticle Hamiltonian
$\tilde{H}\left(\tilde{\epsilon}_{0},\tilde{v}_{0}\right)$ of our
problem which we use to calculate the zero temperature conductance.
The occupancy of the dots and its fluctuations are calculated from the
ground state wavefunction $\left|0\right\rangle $.

\begin{figure}[htbp]
\centering
\includegraphics[clip,width=8cm]{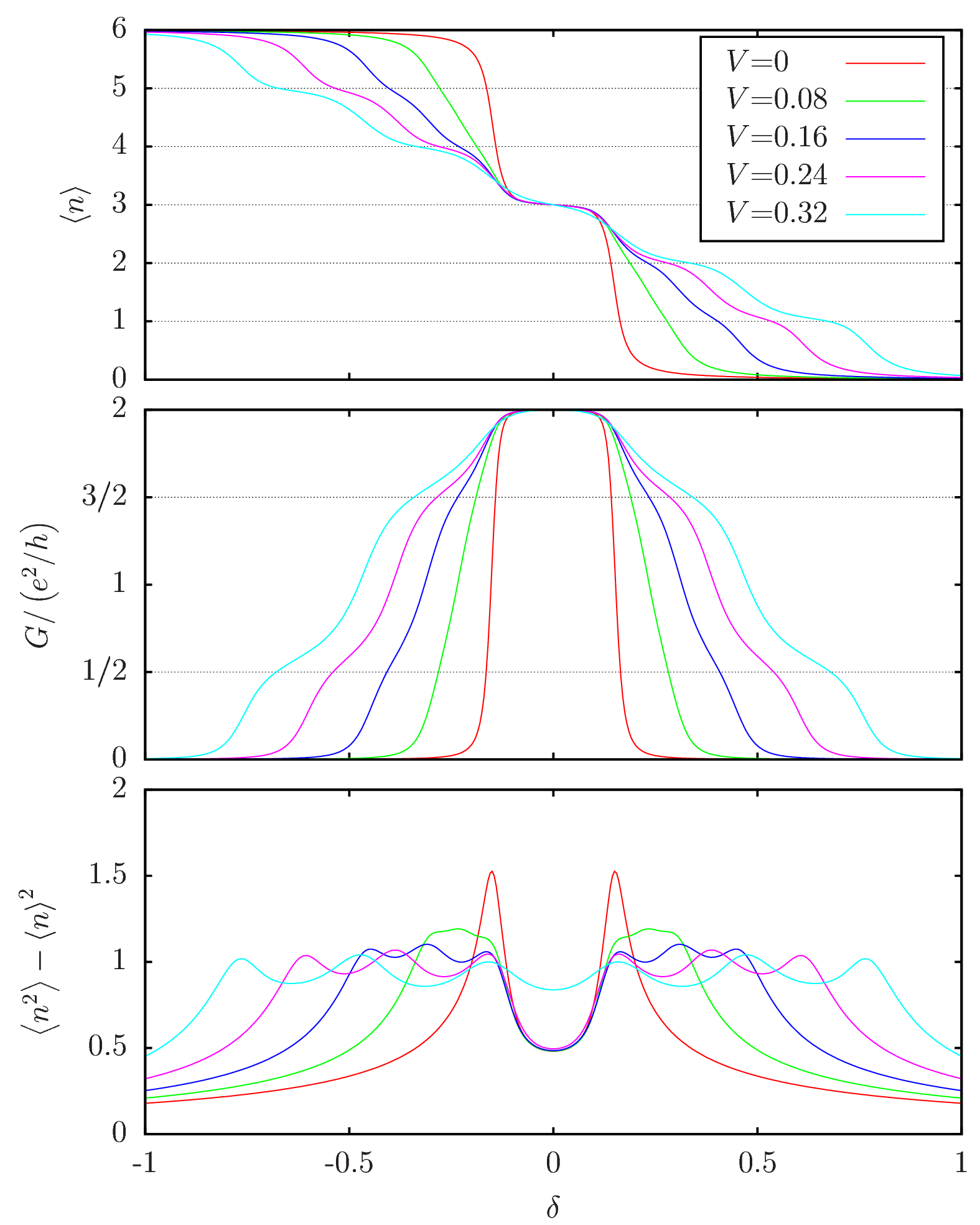}
\caption{(Color online) Occupancy $n$, differential conductance $G$
and charge fluctuations $\delta n^2=\langle n^2\rangle-\langle
n\rangle^2$ as a function of the on-site energy
$\epsilon=-U/2-2V+\delta$ for on-site repulsion $U=0.32$
and for a range of interdot repulsions $V$.
$\Gamma=0.01$.} \label{fig15}
\end{figure}

To study the interplay of the inter-site and on-site charge repulsion
terms $U$ and $V$, we first fix $U$ and increase $V$. For $V=0$, the
system consists simply of three copies of the single-impurity Anderson
model, the properties of which are well known \cite{krishna1980a,
krishna1980b}. In the interval $-U+\Gamma \lesssim \epsilon \lesssim
-\Gamma$, we expect the emergence of the SU(2) Kondo effect in each of
the three channels independently, thus a plateau of unitary
conductance through each dot (note that the unitary limit is now
$2e^2/h$ due to the spin factor). For non-zero but moderate $V<U$, we
expect the occurrence of collective Kondo screening which affects all
three quantum dots. In this case, in addition to the spin degree of
freedom on each dot, there is an orbital degree of freedom, as in the
spinless case that has been discussed in the previous sections. The
two degrees of freedom may become intertwined like in the $\SU(4)$
Kondo effect in carbon nanotubes where the spin and isospin degrees of
freedom are combined. Here, however, the symmetry is
$\SU(2)_\mathrm{spin} \times \SU(3)_\mathrm{orbital}$ and richer
behavior is expected. This model is closely related to the impurity
models studied in the context of dynamical mean-field theory for
correlated bulk metals, in particular for transition-metal compounds
where the three-fold degenerate $t_{2g}$ $d$-electron orbitals play
the main role. The results in Fig.~\ref{fig15} indicate the emergence
of three different kinds of the Kondo plateau for on-site energies
corresponding to the total TQD occupancy being pinned to an integer
value. At the p-h symmetric point, we observe a Kondo plateau for all
values of $V \leq U$; this is thus contrary to the behavior found in
the spinless model where in this regimes the valence fluctuates and
the Kondo effect does not occur. Two additional types of the Kondo
effect are present, one for single-electron (single-hole) occupancy,
$\langle n \rangle=1,5$, characterized by the conductance pinned to
$G=2e^2/h\sin^2[(\pi/2)(1/3)]=0.5 (e^2/h)$, another for two-electron
(two-hole) occupancy, $\langle n \rangle=2,4$, characterized by
$G=2e^2/h \sin^2[(\pi/2)(2/3)]=1.5 (e^2/h)$. In the first case, the
single electron has both spin and orbital degree of freedom (both in
the fundamental $\SU(2)$ and $\SU(3)$ representations), thus this
corresponds to the $\SU(6)$ Kondo effect. In the second case, two
electrons are comined into a more complex object which is then
Kondo screened.

\section{Conclusions}

We have demonstrated the formation of the $\SU(3)$ singlet Kondo state
in a spinless triple quantum dot system.  We have performed a
thermodynamics analysis with the help of the magnetic susceptibility
($\chi_\mathrm{imp}$) and the impurity entropy ($S_\mathrm{imp}$). For
a highly symmetric triple quantum dot we find, when the dot system
charge is either $n=2$ or $n=1$ (away from the particle-hole symmetric
point for which $n=1.5$), that the system evolves from the
free-orbital regime at high temperatures with $T\chi_\mathrm{imp}=1$
and $S_\mathrm{imp}=\ln 8$ to the three-fold degenerate local moment
fixed point with $T\chi_\mathrm{imp}=4/3$ and $S_\mathrm{imp}=\ln 3$
towards a non-degenerate strong coupling fixed point where the
$\SU(3)$ spin is totally screened. This cross-over occurs on an
exponentially low temperature scale, $\ln T_K \propto -1/\Gamma$. In
contrast, for the electron-hole symmetric point,
the system evolves from the free-local moment regime to a
valence-fluctuating fixed point where the entropy $S_\mathrm{imp}=\ln
6$, and then to zero on the scale of $\Gamma$.  Additionally, we have
investigated possible perturbations that affect the formation of the
$\SU(3)$ singlet Kondo state in real setups. Among these perturbations
we have studied how non-symmetrical lead-dot couplings or on-site
potential energies, and finite inter-dot tunneling rates break the
$\SU(3)$ Kondo physics. We have described the procedure to restore the
$\SU(3)$ Kondo physics in real systems by properly tuning the on-site
dot energies. Finally, we described the more complex behavior of the
spinfull case where the Kondo effect occurs for all integer
occupancies of the triple quantum dot, including at the particle-hole
symmetric point.

\begin{acknowledgments}
R. L. was supported by Spanish MICINN (Grant No. FIS2008-00781, and
FIS2011-23526). T. R. and R. \v{Z}. acknowledge the support of the
Slovenian Research Agency (ARRS) under Program No. P1-0044. 
\end{acknowledgments}


\begin{thebibliography}{63}
\expandafter\ifx\csname natexlab\endcsname\relax\def\natexlab#1{#1}\fi
\expandafter\ifx\csname bibnamefont\endcsname\relax
  \def\bibnamefont#1{#1}\fi
\expandafter\ifx\csname bibfnamefont\endcsname\relax
  \def\bibfnamefont#1{#1}\fi
\expandafter\ifx\csname citenamefont\endcsname\relax
  \def\citenamefont#1{#1}\fi
\expandafter\ifx\csname url\endcsname\relax
  \def\url#1{\texttt{#1}}\fi
\expandafter\ifx\csname urlprefix\endcsname\relax\def\urlprefix{URL }\fi
\providecommand{\bibinfo}[2]{#2}
\providecommand{\eprint}[2][]{\url{#2}}

\bibitem[{\citenamefont{Hewson}(1993)}]{hewson1993}
\bibinfo{author}{\bibfnamefont{A.~C.} \bibnamefont{Hewson}},
  \bibinfo{journal}{Phys. Rev. Lett.} \textbf{\bibinfo{volume}{70}},
  \bibinfo{pages}{4007} (\bibinfo{year}{1993}).

\bibitem[{\citenamefont{Ng and Lee}(1988)}]{ng1988}
\bibinfo{author}{\bibfnamefont{T.~K.} \bibnamefont{Ng}} \bibnamefont{and}
  \bibinfo{author}{\bibfnamefont{P.~A.} \bibnamefont{Lee}},
  \bibinfo{journal}{Phys. Rev. Lett.} \textbf{\bibinfo{volume}{61}},
  \bibinfo{pages}{1768} (\bibinfo{year}{1988}).

\bibitem[{\citenamefont{Glazman and Raikh}(1988)}]{glazman1988}
\bibinfo{author}{\bibfnamefont{L.~I.} \bibnamefont{Glazman}} \bibnamefont{and}
  \bibinfo{author}{\bibfnamefont{M.~E.} \bibnamefont{Raikh}},
  \bibinfo{journal}{JETP Lett.} \textbf{\bibinfo{volume}{47}},
  \bibinfo{pages}{452} (\bibinfo{year}{1988}).

\bibitem[{\citenamefont{Meir et~al.}(1993)\citenamefont{Meir, Wingreen, and
  Lee}}]{meir1993}
\bibinfo{author}{\bibfnamefont{Y.}~\bibnamefont{Meir}},
  \bibinfo{author}{\bibfnamefont{N.~S.} \bibnamefont{Wingreen}},
  \bibnamefont{and} \bibinfo{author}{\bibfnamefont{P.~A.} \bibnamefont{Lee}},
  \bibinfo{journal}{Phys. Rev. Lett.} \textbf{\bibinfo{volume}{70}},
  \bibinfo{pages}{2601} (\bibinfo{year}{1993}).

\bibitem[{\citenamefont{Goldhaber-Gordon
  et~al.}(1998{\natexlab{a}})\citenamefont{Goldhaber-Gordon, Shtrikman, Mahalu,
  Abusch-Magder, Meirav, and Kastner}}]{goldhabergordon1998b}
\bibinfo{author}{\bibfnamefont{D.}~\bibnamefont{Goldhaber-Gordon}},
  \bibinfo{author}{\bibfnamefont{H.}~\bibnamefont{Shtrikman}},
  \bibinfo{author}{\bibfnamefont{D.}~\bibnamefont{Mahalu}},
  \bibinfo{author}{\bibfnamefont{D.}~\bibnamefont{Abusch-Magder}},
  \bibinfo{author}{\bibfnamefont{U.}~\bibnamefont{Meirav}}, \bibnamefont{and}
  \bibinfo{author}{\bibfnamefont{M.~A.} \bibnamefont{Kastner}},
  \bibinfo{journal}{Nature} \textbf{\bibinfo{volume}{391}},
  \bibinfo{pages}{156} (\bibinfo{year}{1998}{\natexlab{a}}).

\bibitem[{\citenamefont{Cronenwett et~al.}(1998)\citenamefont{Cronenwett,
  Oosterkamp, and Kouwenhoven}}]{cronenwett1998}
\bibinfo{author}{\bibfnamefont{S.~M.} \bibnamefont{Cronenwett}},
  \bibinfo{author}{\bibfnamefont{T.~H.} \bibnamefont{Oosterkamp}},
  \bibnamefont{and} \bibinfo{author}{\bibfnamefont{L.~P.}
  \bibnamefont{Kouwenhoven}}, \bibinfo{journal}{Science}
  \textbf{\bibinfo{volume}{281}}, \bibinfo{pages}{540} (\bibinfo{year}{1998}).

\bibitem[{\citenamefont{van~der Wiel et~al.}(2000)\citenamefont{van~der Wiel,
  Franceschi, Fujisawa, Elzerman, Tarucha, and Kouwenhoven}}]{wiel2000}
\bibinfo{author}{\bibfnamefont{W.~G.} \bibnamefont{van~der Wiel}},
  \bibinfo{author}{\bibfnamefont{S.~D.} \bibnamefont{Franceschi}},
  \bibinfo{author}{\bibfnamefont{T.}~\bibnamefont{Fujisawa}},
  \bibinfo{author}{\bibfnamefont{J.~M.} \bibnamefont{Elzerman}},
  \bibinfo{author}{\bibfnamefont{S.}~\bibnamefont{Tarucha}}, \bibnamefont{and}
  \bibinfo{author}{\bibfnamefont{L.~P.} \bibnamefont{Kouwenhoven}},
  \bibinfo{journal}{Science} \textbf{\bibinfo{volume}{289}},
  \bibinfo{pages}{2105} (\bibinfo{year}{2000}).

\bibitem[{\citenamefont{Kouwenhoven et~al.}(1997)\citenamefont{Kouwenhoven,
  Oosterkamp, Danoesastro, Eto, Austing, Honda, and Tarucha}}]{kouwenhoven1997}
\bibinfo{author}{\bibfnamefont{L.~P.} \bibnamefont{Kouwenhoven}},
  \bibinfo{author}{\bibfnamefont{T.~H.} \bibnamefont{Oosterkamp}},
  \bibinfo{author}{\bibfnamefont{M.~W.~S.} \bibnamefont{Danoesastro}},
  \bibinfo{author}{\bibfnamefont{M.}~\bibnamefont{Eto}},
  \bibinfo{author}{\bibfnamefont{D.~G.} \bibnamefont{Austing}},
  \bibinfo{author}{\bibfnamefont{T.}~\bibnamefont{Honda}}, \bibnamefont{and}
  \bibinfo{author}{\bibfnamefont{S.}~\bibnamefont{Tarucha}},
  \bibinfo{journal}{Science} \textbf{\bibinfo{volume}{278}},
  \bibinfo{pages}{1788} (\bibinfo{year}{1997}).

\bibitem[{\citenamefont{Schmid et~al.}(1998)\citenamefont{Schmid, Weis, Eberl,
  and v.~Klitzing}}]{Schmid1998182}
\bibinfo{author}{\bibfnamefont{J.}~\bibnamefont{Schmid}},
  \bibinfo{author}{\bibfnamefont{J.}~\bibnamefont{Weis}},
  \bibinfo{author}{\bibfnamefont{K.}~\bibnamefont{Eberl}}, \bibnamefont{and}
  \bibinfo{author}{\bibfnamefont{K.}~\bibnamefont{v.~Klitzing}},
  \bibinfo{journal}{Physica B: Condensed Matter}
  \textbf{\bibinfo{volume}{256‚Äö√Ñ√¨258}}, \bibinfo{pages}{182 }
  (\bibinfo{year}{1998}).

\bibitem[{\citenamefont{Ralph and Buhrman}(1994)}]{ralph1994equib}
\bibinfo{author}{\bibfnamefont{D.~C.} \bibnamefont{Ralph}} \bibnamefont{and}
  \bibinfo{author}{\bibfnamefont{R.~A.} \bibnamefont{Buhrman}},
  \bibinfo{journal}{Phys. Rev. Lett.} \textbf{\bibinfo{volume}{72}},
  \bibinfo{pages}{3401} (\bibinfo{year}{1994}).

\bibitem[{\citenamefont{Kogan et~al.}(2004)\citenamefont{Kogan, Amasha, and
  Kastner}}]{Kogan28052004}
\bibinfo{author}{\bibfnamefont{A.}~\bibnamefont{Kogan}},
  \bibinfo{author}{\bibfnamefont{S.}~\bibnamefont{Amasha}}, \bibnamefont{and}
  \bibinfo{author}{\bibfnamefont{M.~A.} \bibnamefont{Kastner}},
  \bibinfo{journal}{Science} \textbf{\bibinfo{volume}{304}},
  \bibinfo{pages}{1293} (\bibinfo{year}{2004}).

\bibitem[{\citenamefont{De~Franceschi et~al.}(2002)\citenamefont{De~Franceschi,
  Hanson, van~der Wiel, Elzerman, Wijpkema, Fujisawa, Tarucha, and
  Kouwenhoven}}]{PhysRevLett.89.156801}
\bibinfo{author}{\bibfnamefont{S.}~\bibnamefont{De~Franceschi}},
  \bibinfo{author}{\bibfnamefont{R.}~\bibnamefont{Hanson}},
  \bibinfo{author}{\bibfnamefont{W.~G.} \bibnamefont{van~der Wiel}},
  \bibinfo{author}{\bibfnamefont{J.~M.} \bibnamefont{Elzerman}},
  \bibinfo{author}{\bibfnamefont{J.~J.} \bibnamefont{Wijpkema}},
  \bibinfo{author}{\bibfnamefont{T.}~\bibnamefont{Fujisawa}},
  \bibinfo{author}{\bibfnamefont{S.}~\bibnamefont{Tarucha}}, \bibnamefont{and}
  \bibinfo{author}{\bibfnamefont{L.~P.} \bibnamefont{Kouwenhoven}},
  \bibinfo{journal}{Phys. Rev. Lett.} \textbf{\bibinfo{volume}{89}},
  \bibinfo{pages}{156801} (\bibinfo{year}{2002}).

\bibitem[{\citenamefont{S\'anchez and L\'opez}(2005)}]{PhysRevB.71.035315}
\bibinfo{author}{\bibfnamefont{D.}~\bibnamefont{S\'anchez}} \bibnamefont{and}
  \bibinfo{author}{\bibfnamefont{R.}~\bibnamefont{L\'opez}},
  \bibinfo{journal}{Phys. Rev. B} \textbf{\bibinfo{volume}{71}},
  \bibinfo{pages}{035315} (\bibinfo{year}{2005}).

\bibitem[{\citenamefont{Aguado and Langreth}(2000)}]{aguado2000}
\bibinfo{author}{\bibfnamefont{R.}~\bibnamefont{Aguado}} \bibnamefont{and}
  \bibinfo{author}{\bibfnamefont{D.~C.} \bibnamefont{Langreth}},
  \bibinfo{journal}{Phys. Rev. Lett.} \textbf{\bibinfo{volume}{85}},
  \bibinfo{pages}{1946} (\bibinfo{year}{2000}).

\bibitem[{\citenamefont{Fuhrer et~al.}(2003)\citenamefont{Fuhrer, Ihn, Ensslin,
  Wegscheider, and Bichler}}]{fuhrer2004b}
\bibinfo{author}{\bibfnamefont{A.}~\bibnamefont{Fuhrer}},
  \bibinfo{author}{\bibfnamefont{T.}~\bibnamefont{Ihn}},
  \bibinfo{author}{\bibfnamefont{K.}~\bibnamefont{Ensslin}},
  \bibinfo{author}{\bibfnamefont{W.}~\bibnamefont{Wegscheider}},
  \bibnamefont{and} \bibinfo{author}{\bibfnamefont{M.}~\bibnamefont{Bichler}},
  \bibinfo{journal}{Phys. Rev. Lett.} \textbf{\bibinfo{volume}{93}},
  \bibinfo{pages}{176803} (\bibinfo{year}{2003}).

\bibitem[{\citenamefont{Hofstetter and Zar\'and}(2004)}]{hofstetter2004}
\bibinfo{author}{\bibfnamefont{W.}~\bibnamefont{Hofstetter}} \bibnamefont{and}
  \bibinfo{author}{\bibfnamefont{G.}~\bibnamefont{Zar\'and}},
  \bibinfo{journal}{Phys. Rev. B} \textbf{\bibinfo{volume}{69}},
  \bibinfo{pages}{235301} (\bibinfo{year}{2004}).

\bibitem[{\citenamefont{Kogan et~al.}(2003)\citenamefont{Kogan, Granger,
  Kastner, Goldhaber-Gordon, and Shtrikman}}]{kogan2003}
\bibinfo{author}{\bibfnamefont{A.}~\bibnamefont{Kogan}},
  \bibinfo{author}{\bibfnamefont{G.}~\bibnamefont{Granger}},
  \bibinfo{author}{\bibfnamefont{M.~A.} \bibnamefont{Kastner}},
  \bibinfo{author}{\bibfnamefont{D.}~\bibnamefont{Goldhaber-Gordon}},
  \bibnamefont{and}
  \bibinfo{author}{\bibfnamefont{H.}~\bibnamefont{Shtrikman}},
  \bibinfo{journal}{Phys. Rev. B} \textbf{\bibinfo{volume}{67}},
  \bibinfo{pages}{113309} (\bibinfo{year}{2003}).

\bibitem[{\citenamefont{Pustilnik et~al.}(2003)\citenamefont{Pustilnik,
  Glazman, and Hofstetter}}]{pustilnik2003}
\bibinfo{author}{\bibfnamefont{M.}~\bibnamefont{Pustilnik}},
  \bibinfo{author}{\bibfnamefont{L.~I.} \bibnamefont{Glazman}},
  \bibnamefont{and}
  \bibinfo{author}{\bibfnamefont{W.}~\bibnamefont{Hofstetter}},
  \bibinfo{journal}{Phys. Rev. B} \textbf{\bibinfo{volume}{68}},
  \bibinfo{pages}{161303(R)} (\bibinfo{year}{2003}).

\bibitem[{\citenamefont{Pustilnik and Glazman}(2001)}]{pustilnik2001st}
\bibinfo{author}{\bibfnamefont{M.}~\bibnamefont{Pustilnik}} \bibnamefont{and}
  \bibinfo{author}{\bibfnamefont{L.~I.} \bibnamefont{Glazman}},
  \bibinfo{journal}{Phys. Rev. B} \textbf{\bibinfo{volume}{64}},
  \bibinfo{pages}{045328} (\bibinfo{year}{2001}).

\bibitem[{\citenamefont{Sasaki et~al.}(2000)\citenamefont{Sasaki,
  de~Franceschi, Elzerman, van~der Wiel, Eto, Tarucha, and
  Kouwenhoven}}]{sasaki2000}
\bibinfo{author}{\bibfnamefont{S.}~\bibnamefont{Sasaki}},
  \bibinfo{author}{\bibfnamefont{S.}~\bibnamefont{de~Franceschi}},
  \bibinfo{author}{\bibfnamefont{J.~M.} \bibnamefont{Elzerman}},
  \bibinfo{author}{\bibfnamefont{W.~G.} \bibnamefont{van~der Wiel}},
  \bibinfo{author}{\bibfnamefont{M.}~\bibnamefont{Eto}},
  \bibinfo{author}{\bibfnamefont{S.}~\bibnamefont{Tarucha}}, \bibnamefont{and}
  \bibinfo{author}{\bibfnamefont{L.~P.} \bibnamefont{Kouwenhoven}},
  \bibinfo{journal}{Nature} \textbf{\bibinfo{volume}{405}},
  \bibinfo{pages}{764} (\bibinfo{year}{2000}).

\bibitem[{\citenamefont{Nygard et~al.}(2000)\citenamefont{Nygard, Cobden, and
  Lindelof}}]{nygard2000}
\bibinfo{author}{\bibfnamefont{J.}~\bibnamefont{Nygard}},
  \bibinfo{author}{\bibfnamefont{D.~H.} \bibnamefont{Cobden}},
  \bibnamefont{and} \bibinfo{author}{\bibfnamefont{P.~E.}
  \bibnamefont{Lindelof}}, \bibinfo{journal}{Nature}
  \textbf{\bibinfo{volume}{408}}, \bibinfo{pages}{342} (\bibinfo{year}{2000}).

\bibitem[{\citenamefont{Odom et~al.}(2000)\citenamefont{Odom, Huang, Cheung,
  and Lieber}}]{odom2000}
\bibinfo{author}{\bibfnamefont{T.~W.} \bibnamefont{Odom}},
  \bibinfo{author}{\bibfnamefont{J.-L.} \bibnamefont{Huang}},
  \bibinfo{author}{\bibfnamefont{C.~L.} \bibnamefont{Cheung}},
  \bibnamefont{and} \bibinfo{author}{\bibfnamefont{C.~M.}
  \bibnamefont{Lieber}}, \bibinfo{journal}{Science}
  \textbf{\bibinfo{volume}{290}}, \bibinfo{pages}{1549} (\bibinfo{year}{2000}).

\bibitem[{\citenamefont{Jarillo-Herrero
  et~al.}(2005)\citenamefont{Jarillo-Herrero, Kong, S.J van~der Zant, Dekker,
  and Kouwenhoven}}]{jarillo}
\bibinfo{author}{\bibfnamefont{P.}~\bibnamefont{Jarillo-Herrero}},
  \bibinfo{author}{\bibfnamefont{J.}~\bibnamefont{Kong}},
  \bibinfo{author}{\bibfnamefont{H.}~\bibnamefont{S.J van~der Zant}},
  \bibinfo{author}{\bibfnamefont{C.}~\bibnamefont{Dekker}}, \bibnamefont{and}
  \bibinfo{author}{\bibfnamefont{L.}~\bibnamefont{Kouwenhoven}},
  \bibinfo{journal}{Nature} \textbf{\bibinfo{volume}{434}},
  \bibinfo{pages}{484} (\bibinfo{year}{2005}).

\bibitem[{\citenamefont{Lim et~al.}(2006)\citenamefont{Lim, Choi, Choi,
  L\'opez, and Aguado}}]{PhysRevB.74.205119}
\bibinfo{author}{\bibfnamefont{J.~S.} \bibnamefont{Lim}},
  \bibinfo{author}{\bibfnamefont{M.-S.} \bibnamefont{Choi}},
  \bibinfo{author}{\bibfnamefont{M.~Y.} \bibnamefont{Choi}},
  \bibinfo{author}{\bibfnamefont{R.}~\bibnamefont{L\'opez}}, \bibnamefont{and}
  \bibinfo{author}{\bibfnamefont{R.}~\bibnamefont{Aguado}},
  \bibinfo{journal}{Phys. Rev. B} \textbf{\bibinfo{volume}{74}},
  \bibinfo{pages}{205119} (\bibinfo{year}{2006}).

\bibitem[{\citenamefont{Makarovski et~al.}(2007)\citenamefont{Makarovski, Liu,
  and Finkelstein}}]{PhysRevLett.99.066801}
\bibinfo{author}{\bibfnamefont{A.}~\bibnamefont{Makarovski}},
  \bibinfo{author}{\bibfnamefont{J.}~\bibnamefont{Liu}}, \bibnamefont{and}
  \bibinfo{author}{\bibfnamefont{G.}~\bibnamefont{Finkelstein}},
  \bibinfo{journal}{Phys. Rev. Lett.} \textbf{\bibinfo{volume}{99}},
  \bibinfo{pages}{066801} (\bibinfo{year}{2007}).

\bibitem[{\citenamefont{Jacob and Kotliar}(2010)}]{PhysRevB.82.085423}
\bibinfo{author}{\bibfnamefont{D.}~\bibnamefont{Jacob}} \bibnamefont{and}
  \bibinfo{author}{\bibfnamefont{G.}~\bibnamefont{Kotliar}},
  \bibinfo{journal}{Phys. Rev. B} \textbf{\bibinfo{volume}{82}},
  \bibinfo{pages}{085423} (\bibinfo{year}{2010}).

\bibitem[{\citenamefont{B\"usser et~al.}(2011)\citenamefont{B\"usser, Vernek,
  Orellana, Lara, Kim, Feiguin, Anda, and Martins}}]{PhysRevB.83.125404}
\bibinfo{author}{\bibfnamefont{C.~A.} \bibnamefont{B\"usser}},
  \bibinfo{author}{\bibfnamefont{E.}~\bibnamefont{Vernek}},
  \bibinfo{author}{\bibfnamefont{P.}~\bibnamefont{Orellana}},
  \bibinfo{author}{\bibfnamefont{G.~A.} \bibnamefont{Lara}},
  \bibinfo{author}{\bibfnamefont{E.~H.} \bibnamefont{Kim}},
  \bibinfo{author}{\bibfnamefont{A.~E.} \bibnamefont{Feiguin}},
  \bibinfo{author}{\bibfnamefont{E.~V.} \bibnamefont{Anda}}, \bibnamefont{and}
  \bibinfo{author}{\bibfnamefont{G.~B.} \bibnamefont{Martins}},
  \bibinfo{journal}{Phys. Rev. B} \textbf{\bibinfo{volume}{83}},
  \bibinfo{pages}{125404} (\bibinfo{year}{2011}).

\bibitem[{\citenamefont{Cornaglia et~al.}(2009)\citenamefont{Cornaglia, Usaj,
  and Balseiro}}]{PhysRevLett.102.046801}
\bibinfo{author}{\bibfnamefont{P.~S.} \bibnamefont{Cornaglia}},
  \bibinfo{author}{\bibfnamefont{G.}~\bibnamefont{Usaj}}, \bibnamefont{and}
  \bibinfo{author}{\bibfnamefont{C.~A.} \bibnamefont{Balseiro}},
  \bibinfo{journal}{Phys. Rev. Lett.} \textbf{\bibinfo{volume}{102}},
  \bibinfo{pages}{046801} (\bibinfo{year}{2009}).

\bibitem[{\citenamefont{Anders et~al.}(2008)\citenamefont{Anders, Logan,
  Galpin, and Finkelstein}}]{PhysRevLett.100.086809}
\bibinfo{author}{\bibfnamefont{F.~B.} \bibnamefont{Anders}},
  \bibinfo{author}{\bibfnamefont{D.~E.} \bibnamefont{Logan}},
  \bibinfo{author}{\bibfnamefont{M.~R.} \bibnamefont{Galpin}},
  \bibnamefont{and}
  \bibinfo{author}{\bibfnamefont{G.}~\bibnamefont{Finkelstein}},
  \bibinfo{journal}{Phys. Rev. Lett.} \textbf{\bibinfo{volume}{100}},
  \bibinfo{pages}{086809} (\bibinfo{year}{2008}).

\bibitem[{\citenamefont{Lim et~al.}(2011)\citenamefont{Lim, L\'opez, Giorgi,
  and S\'anchez}}]{PhysRevB.83.155325}
\bibinfo{author}{\bibfnamefont{J.~S.} \bibnamefont{Lim}},
  \bibinfo{author}{\bibfnamefont{R.}~\bibnamefont{L\'opez}},
  \bibinfo{author}{\bibfnamefont{G.~L.} \bibnamefont{Giorgi}},
  \bibnamefont{and}
  \bibinfo{author}{\bibfnamefont{D.}~\bibnamefont{S\'anchez}},
  \bibinfo{journal}{Phys. Rev. B} \textbf{\bibinfo{volume}{83}},
  \bibinfo{pages}{155325} (\bibinfo{year}{2011}).

\bibitem[{\citenamefont{Klochan et~al.}(2011)\citenamefont{Klochan, Micolich,
  Hamilton, Trunov, Reuter, and Wieck}}]{PhysRevLett.107.076805}
\bibinfo{author}{\bibfnamefont{O.}~\bibnamefont{Klochan}},
  \bibinfo{author}{\bibfnamefont{A.~P.} \bibnamefont{Micolich}},
  \bibinfo{author}{\bibfnamefont{A.~R.} \bibnamefont{Hamilton}},
  \bibinfo{author}{\bibfnamefont{K.}~\bibnamefont{Trunov}},
  \bibinfo{author}{\bibfnamefont{D.}~\bibnamefont{Reuter}}, \bibnamefont{and}
  \bibinfo{author}{\bibfnamefont{A.~D.} \bibnamefont{Wieck}},
  \bibinfo{journal}{Phys. Rev. Lett.} \textbf{\bibinfo{volume}{107}},
  \bibinfo{pages}{076805} (\bibinfo{year}{2011}).

\bibitem[{\citenamefont{Borda et~al.}(2003)\citenamefont{Borda, Zar\'and,
  Hofstetter, Halperin, and von Delft}}]{PhysRevLett.90.026602}
\bibinfo{author}{\bibfnamefont{L.}~\bibnamefont{Borda}},
  \bibinfo{author}{\bibfnamefont{G.}~\bibnamefont{Zar\'and}},
  \bibinfo{author}{\bibfnamefont{W.}~\bibnamefont{Hofstetter}},
  \bibinfo{author}{\bibfnamefont{B.~I.} \bibnamefont{Halperin}},
  \bibnamefont{and} \bibinfo{author}{\bibfnamefont{J.}~\bibnamefont{von
  Delft}}, \bibinfo{journal}{Phys. Rev. Lett.} \textbf{\bibinfo{volume}{90}},
  \bibinfo{pages}{026602} (\bibinfo{year}{2003}).

\bibitem[{\citenamefont{L\'opez et~al.}(2005)\citenamefont{L\'opez, S\'anchez,
  Lee, Choi, Simon, and Le~Hur}}]{PhysRevB.71.115312}
\bibinfo{author}{\bibfnamefont{R.}~\bibnamefont{L\'opez}},
  \bibinfo{author}{\bibfnamefont{D.}~\bibnamefont{S\'anchez}},
  \bibinfo{author}{\bibfnamefont{M.}~\bibnamefont{Lee}},
  \bibinfo{author}{\bibfnamefont{M.-S.} \bibnamefont{Choi}},
  \bibinfo{author}{\bibfnamefont{P.}~\bibnamefont{Simon}}, \bibnamefont{and}
  \bibinfo{author}{\bibfnamefont{K.}~\bibnamefont{Le~Hur}},
  \bibinfo{journal}{Phys. Rev. B} \textbf{\bibinfo{volume}{71}},
  \bibinfo{pages}{115312} (\bibinfo{year}{2005}).

\bibitem[{\citenamefont{Sasaki et~al.}(2004)\citenamefont{Sasaki, Amaha,
  Asakawa, Eto, and Tarucha}}]{PhysRevLett.93.017205}
\bibinfo{author}{\bibfnamefont{S.}~\bibnamefont{Sasaki}},
  \bibinfo{author}{\bibfnamefont{S.}~\bibnamefont{Amaha}},
  \bibinfo{author}{\bibfnamefont{N.}~\bibnamefont{Asakawa}},
  \bibinfo{author}{\bibfnamefont{M.}~\bibnamefont{Eto}}, \bibnamefont{and}
  \bibinfo{author}{\bibfnamefont{S.}~\bibnamefont{Tarucha}},
  \bibinfo{journal}{Phys. Rev. Lett.} \textbf{\bibinfo{volume}{93}},
  \bibinfo{pages}{017205} (\bibinfo{year}{2004}).

\bibitem[{\citenamefont{Liang et~al.}(2002)\citenamefont{Liang, Bockrath, and
  Park}}]{PhysRevLett.88.126801}
\bibinfo{author}{\bibfnamefont{W.}~\bibnamefont{Liang}},
  \bibinfo{author}{\bibfnamefont{M.}~\bibnamefont{Bockrath}}, \bibnamefont{and}
  \bibinfo{author}{\bibfnamefont{H.}~\bibnamefont{Park}},
  \bibinfo{journal}{Phys. Rev. Lett.} \textbf{\bibinfo{volume}{88}},
  \bibinfo{pages}{126801} (\bibinfo{year}{2002}).

\bibitem[{\citenamefont{Cobden and Nyg\aa{}rd}(2002)}]{PhysRevLett.89.046803}
\bibinfo{author}{\bibfnamefont{D.~H.} \bibnamefont{Cobden}} \bibnamefont{and}
  \bibinfo{author}{\bibfnamefont{J.}~\bibnamefont{Nyg\aa{}rd}},
  \bibinfo{journal}{Phys. Rev. Lett.} \textbf{\bibinfo{volume}{89}},
  \bibinfo{pages}{046803} (\bibinfo{year}{2002}).

\bibitem[{\citenamefont{Choi et~al.}(2005)\citenamefont{Choi, L\'opez, and
  Aguado}}]{PhysRevLett.95.067204}
\bibinfo{author}{\bibfnamefont{M.-S.} \bibnamefont{Choi}},
  \bibinfo{author}{\bibfnamefont{R.}~\bibnamefont{L\'opez}}, \bibnamefont{and}
  \bibinfo{author}{\bibfnamefont{R.}~\bibnamefont{Aguado}},
  \bibinfo{journal}{Phys. Rev. Lett.} \textbf{\bibinfo{volume}{95}},
  \bibinfo{pages}{067204} (\bibinfo{year}{2005}).

\bibitem[{\citenamefont{Amasha et~al.}(2012)\citenamefont{Amasha, Keller, Rau,
  Carmi, Katine, Shtrikman, Oreg, and Goldhaber-Gordon}}]{goldhaber2012}
\bibinfo{author}{\bibfnamefont{S.}~\bibnamefont{Amasha}},
  \bibinfo{author}{\bibfnamefont{J.}~\bibnamefont{Keller}},
  \bibinfo{author}{\bibfnamefont{I.}~\bibnamefont{Rau}},
  \bibinfo{author}{\bibfnamefont{A.}~\bibnamefont{Carmi}},
  \bibinfo{author}{\bibfnamefont{J.}~\bibnamefont{Katine}},
  \bibinfo{author}{\bibfnamefont{H.}~\bibnamefont{Shtrikman}},
  \bibinfo{author}{\bibfnamefont{Y.}~\bibnamefont{Oreg}}, \bibnamefont{and}
  \bibinfo{author}{\bibfnamefont{D.}~\bibnamefont{Goldhaber-Gordon}},
  \bibinfo{howpublished}{arxiv: 1207.0526},
  (\bibinfo{year}{2012}).

\bibitem[{\citenamefont{Carmi et~al.}(2011)\citenamefont{Carmi, Oreg, and
  Berkooz}}]{PhysRevLett.106.106401}
\bibinfo{author}{\bibfnamefont{A.}~\bibnamefont{Carmi}},
  \bibinfo{author}{\bibfnamefont{Y.}~\bibnamefont{Oreg}}, \bibnamefont{and}
  \bibinfo{author}{\bibfnamefont{M.}~\bibnamefont{Berkooz}},
  \bibinfo{journal}{Phys. Rev. Lett.} \textbf{\bibinfo{volume}{106}},
  \bibinfo{pages}{106401} (\bibinfo{year}{2011}).

\bibitem[{\citenamefont{Mora}(2009)}]{PhysRevB.80.125304}
\bibinfo{author}{\bibfnamefont{C.}~\bibnamefont{Mora}}, \bibinfo{journal}{Phys.
  Rev. B} \textbf{\bibinfo{volume}{80}}, \bibinfo{pages}{125304}
  (\bibinfo{year}{2009}).

\bibitem[{\citenamefont{Mora et~al.}(2009)\citenamefont{Mora, Vitushinsky,
  Leyronas, Clerk, and Le~Hur}}]{PhysRevB.80.155322}
\bibinfo{author}{\bibfnamefont{C.}~\bibnamefont{Mora}},
  \bibinfo{author}{\bibfnamefont{P.}~\bibnamefont{Vitushinsky}},
  \bibinfo{author}{\bibfnamefont{X.}~\bibnamefont{Leyronas}},
  \bibinfo{author}{\bibfnamefont{A.~A.} \bibnamefont{Clerk}}, \bibnamefont{and}
  \bibinfo{author}{\bibfnamefont{K.}~\bibnamefont{Le~Hur}},
  \bibinfo{journal}{Phys. Rev. B} \textbf{\bibinfo{volume}{80}},
  \bibinfo{pages}{155322} (\bibinfo{year}{2009}).

\bibitem[{\citenamefont{Sakano et~al.}(2011)\citenamefont{Sakano, Oguri, Kato,
  and Tarucha}}]{PhysRevB.83.241301}
\bibinfo{author}{\bibfnamefont{R.}~\bibnamefont{Sakano}},
  \bibinfo{author}{\bibfnamefont{A.}~\bibnamefont{Oguri}},
  \bibinfo{author}{\bibfnamefont{T.}~\bibnamefont{Kato}}, \bibnamefont{and}
  \bibinfo{author}{\bibfnamefont{S.}~\bibnamefont{Tarucha}},
  \bibinfo{journal}{Phys. Rev. B} \textbf{\bibinfo{volume}{83}},
  \bibinfo{pages}{241301} (\bibinfo{year}{2011}).

\bibitem[{\citenamefont{Duki}(2011)}]{PhysRevB.83.134423}
\bibinfo{author}{\bibfnamefont{S.~F.} \bibnamefont{Duki}},
  \bibinfo{journal}{Phys. Rev. B} \textbf{\bibinfo{volume}{83}},
  \bibinfo{pages}{134423} (\bibinfo{year}{2011}).

\bibitem[{Ama(2008)}]{Amaha20081322}
\bibinfo{journal}{Physica E: Low-dimensional Systems and Nanostructures}
  \textbf{\bibinfo{volume}{40}}, \bibinfo{pages}{1322} (\bibinfo{year}{2008}).

\bibitem[{\citenamefont{Rogge and Haug}(2008)}]{PhysRevB.77.193306}
\bibinfo{author}{\bibfnamefont{M.~C.} \bibnamefont{Rogge}} \bibnamefont{and}
  \bibinfo{author}{\bibfnamefont{R.~J.} \bibnamefont{Haug}},
  \bibinfo{journal}{Phys. Rev. B} \textbf{\bibinfo{volume}{77}},
  \bibinfo{pages}{193306} (\bibinfo{year}{2008}).

\bibitem[{\citenamefont{Mitchell et~al.}(2009)\citenamefont{Mitchell, Jarrold,
  and Logan}}]{PhysRevB.79.085124}
\bibinfo{author}{\bibfnamefont{A.~K.} \bibnamefont{Mitchell}},
  \bibinfo{author}{\bibfnamefont{T.~F.} \bibnamefont{Jarrold}},
  \bibnamefont{and} \bibinfo{author}{\bibfnamefont{D.~E.} \bibnamefont{Logan}},
  \bibinfo{journal}{Phys. Rev. B} \textbf{\bibinfo{volume}{79}},
  \bibinfo{pages}{085124} (\bibinfo{year}{2009}).

\bibitem[{\citenamefont{Gaudreau et~al.}(2006)\citenamefont{Gaudreau,
  Studenikin, Sachrajda, Zawadzki, Kam, Lapointe, Korkusinski, and
  Hawrylak}}]{PhysRevLett.97.036807}
\bibinfo{author}{\bibfnamefont{L.}~\bibnamefont{Gaudreau}},
  \bibinfo{author}{\bibfnamefont{S.~A.} \bibnamefont{Studenikin}},
  \bibinfo{author}{\bibfnamefont{A.~S.} \bibnamefont{Sachrajda}},
  \bibinfo{author}{\bibfnamefont{P.}~\bibnamefont{Zawadzki}},
  \bibinfo{author}{\bibfnamefont{A.}~\bibnamefont{Kam}},
  \bibinfo{author}{\bibfnamefont{J.}~\bibnamefont{Lapointe}},
  \bibinfo{author}{\bibfnamefont{M.}~\bibnamefont{Korkusinski}},
  \bibnamefont{and} \bibinfo{author}{\bibfnamefont{P.}~\bibnamefont{Hawrylak}},
  \bibinfo{journal}{Phys. Rev. Lett.} \textbf{\bibinfo{volume}{97}},
  \bibinfo{pages}{036807} (\bibinfo{year}{2006}).

\bibitem[{\citenamefont{Granger et~al.}(2010)\citenamefont{Granger, Gaudreau,
  Kam, Pioro-Ladri\`ere, Studenikin, Wasilewski, Zawadzki, and
  Sachrajda}}]{PhysRevB.82.075304}
\bibinfo{author}{\bibfnamefont{G.}~\bibnamefont{Granger}},
  \bibinfo{author}{\bibfnamefont{L.}~\bibnamefont{Gaudreau}},
  \bibinfo{author}{\bibfnamefont{A.}~\bibnamefont{Kam}},
  \bibinfo{author}{\bibfnamefont{M.}~\bibnamefont{Pioro-Ladri\`ere}},
  \bibinfo{author}{\bibfnamefont{S.~A.} \bibnamefont{Studenikin}},
  \bibinfo{author}{\bibfnamefont{Z.~R.} \bibnamefont{Wasilewski}},
  \bibinfo{author}{\bibfnamefont{P.}~\bibnamefont{Zawadzki}}, \bibnamefont{and}
  \bibinfo{author}{\bibfnamefont{A.~S.} \bibnamefont{Sachrajda}},
  \bibinfo{journal}{Phys. Rev. B} \textbf{\bibinfo{volume}{82}},
  \bibinfo{pages}{075304} (\bibinfo{year}{2010}).

\bibitem[{\citenamefont{Raussendorf and Briegel}(2001)}]{PhysRevLett.86.5188}
\bibinfo{author}{\bibfnamefont{R.}~\bibnamefont{Raussendorf}} \bibnamefont{and}
  \bibinfo{author}{\bibfnamefont{H.~J.} \bibnamefont{Briegel}},
  \bibinfo{journal}{Phys. Rev. Lett.} \textbf{\bibinfo{volume}{86}},
  \bibinfo{pages}{5188} (\bibinfo{year}{2001}).

\bibitem[{\citenamefont{Saraga and Loss}(2003)}]{PhysRevLett.90.166803}
\bibinfo{author}{\bibfnamefont{D.~S.} \bibnamefont{Saraga}} \bibnamefont{and}
  \bibinfo{author}{\bibfnamefont{D.}~\bibnamefont{Loss}},
  \bibinfo{journal}{Phys. Rev. Lett.} \textbf{\bibinfo{volume}{90}},
  \bibinfo{pages}{166803} (\bibinfo{year}{2003}).

\bibitem[{\citenamefont{Moca et~al.}(2012)\citenamefont{Moca, Alex, von Delft,
  and Zarand}}]{moca2012}
\bibinfo{author}{\bibfnamefont{C.~P.} \bibnamefont{Moca}},
  \bibinfo{author}{\bibfnamefont{A.}~\bibnamefont{Alex}},
  \bibinfo{author}{\bibfnamefont{J.}~\bibnamefont{von Delft}},
  \bibnamefont{and} \bibinfo{author}{\bibfnamefont{G.}~\bibnamefont{Zarand}},
  \bibinfo{journal}{arXiv:1208.0678}  (\bibinfo{year}{2012}).

\bibitem[{\citenamefont{Zinn-Justin and Andrei}(1998)}]{zinnjustin1998}
\bibinfo{author}{\bibfnamefont{P.}~\bibnamefont{Zinn-Justin}} \bibnamefont{and}
  \bibinfo{author}{\bibfnamefont{N.}~\bibnamefont{Andrei}},
  \bibinfo{journal}{Nucl. Phys. B} \textbf{\bibinfo{volume}{528}},
  \bibinfo{pages}{648} (\bibinfo{year}{1998}).

\bibitem[{\citenamefont{Parcollet et~al.}(1998)\citenamefont{Parcollet,
  Georges, Kotliar, and Sengupta}}]{parcollet1998}
\bibinfo{author}{\bibfnamefont{O.}~\bibnamefont{Parcollet}},
  \bibinfo{author}{\bibfnamefont{A.}~\bibnamefont{Georges}},
  \bibinfo{author}{\bibfnamefont{G.}~\bibnamefont{Kotliar}}, \bibnamefont{and}
  \bibinfo{author}{\bibfnamefont{A.}~\bibnamefont{Sengupta}},
  \bibinfo{journal}{Phys. Rev. B} \textbf{\bibinfo{volume}{58}},
  \bibinfo{pages}{3794} (\bibinfo{year}{1998}).

\bibitem[{\citenamefont{Coqblin and Schrieffer}(1969)}]{coqblin1969}
\bibinfo{author}{\bibfnamefont{B.}~\bibnamefont{Coqblin}} \bibnamefont{and}
  \bibinfo{author}{\bibfnamefont{J.~R.} \bibnamefont{Schrieffer}},
  \bibinfo{journal}{Phys. Rev.} \textbf{\bibinfo{volume}{185}},
  \bibinfo{pages}{847} (\bibinfo{year}{1969}).

\bibitem[{\citenamefont{Krishna-murthy
  et~al.}(1980{\natexlab{a}})\citenamefont{Krishna-murthy, Wilkins, and
  Wilson}}]{krishna1980a}
\bibinfo{author}{\bibfnamefont{H.~R.} \bibnamefont{Krishna-murthy}},
  \bibinfo{author}{\bibfnamefont{J.~W.} \bibnamefont{Wilkins}},
  \bibnamefont{and} \bibinfo{author}{\bibfnamefont{K.~G.}
  \bibnamefont{Wilson}}, \bibinfo{journal}{Phys. Rev. B}
  \textbf{\bibinfo{volume}{21}}, \bibinfo{pages}{1003}
  (\bibinfo{year}{1980}{\natexlab{a}}).

\bibitem[{\citenamefont{Krishna-murthy
  et~al.}(1980{\natexlab{b}})\citenamefont{Krishna-murthy, Wilkins, and
  Wilson}}]{krishna1980b}
\bibinfo{author}{\bibfnamefont{H.~R.} \bibnamefont{Krishna-murthy}},
  \bibinfo{author}{\bibfnamefont{J.~W.} \bibnamefont{Wilkins}},
  \bibnamefont{and} \bibinfo{author}{\bibfnamefont{K.~G.}
  \bibnamefont{Wilson}}, \bibinfo{journal}{Phys. Rev. B}
  \textbf{\bibinfo{volume}{21}}, \bibinfo{pages}{1044}
  (\bibinfo{year}{1980}{\natexlab{b}}).

\bibitem[{\citenamefont{Meir and Wingreen}(1992)}]{meir1992}
\bibinfo{author}{\bibfnamefont{Y.}~\bibnamefont{Meir}} \bibnamefont{and}
  \bibinfo{author}{\bibfnamefont{N.~S.} \bibnamefont{Wingreen}},
  \bibinfo{journal}{Phys. Rev. Lett.} \textbf{\bibinfo{volume}{68}},
  \bibinfo{pages}{2512} (\bibinfo{year}{1992}).

\bibitem[{\citenamefont{Wilson}(1975)}]{wilson1975}
\bibinfo{author}{\bibfnamefont{K.~G.} \bibnamefont{Wilson}},
  \bibinfo{journal}{Rev. Mod. Phys.} \textbf{\bibinfo{volume}{47}},
  \bibinfo{pages}{773} (\bibinfo{year}{1975}).

\bibitem[{\citenamefont{Bulla et~al.}(2008)\citenamefont{Bulla, Costi, and
  Pruschke}}]{bulla2008}
\bibinfo{author}{\bibfnamefont{R.}~\bibnamefont{Bulla}},
  \bibinfo{author}{\bibfnamefont{T.}~\bibnamefont{Costi}}, \bibnamefont{and}
  \bibinfo{author}{\bibfnamefont{T.}~\bibnamefont{Pruschke}},
  \bibinfo{journal}{Rev. Mod. Phys.} \textbf{\bibinfo{volume}{80}},
  \bibinfo{pages}{395} (\bibinfo{year}{2008}).

\bibitem[{\citenamefont{Peters et~al.}(2006)\citenamefont{Peters, Pruschke, and
  Anders}}]{peters2006}
\bibinfo{author}{\bibfnamefont{R.}~\bibnamefont{Peters}},
  \bibinfo{author}{\bibfnamefont{T.}~\bibnamefont{Pruschke}}, \bibnamefont{and}
  \bibinfo{author}{\bibfnamefont{F.~B.} \bibnamefont{Anders}},
  \bibinfo{journal}{Phys. Rev. B} \textbf{\bibinfo{volume}{74}},
  \bibinfo{pages}{245114} (\bibinfo{year}{2006}).

\bibitem[{\citenamefont{Weichselbaum and von Delft}(2007)}]{weichselbaum2007}
\bibinfo{author}{\bibfnamefont{A.}~\bibnamefont{Weichselbaum}}
  \bibnamefont{and} \bibinfo{author}{\bibfnamefont{J.}~\bibnamefont{von
  Delft}}, \bibinfo{journal}{Phys. Rev. Lett.} \textbf{\bibinfo{volume}{99}},
  \bibinfo{pages}{076402} (\bibinfo{year}{2007}).

\bibitem[{\citenamefont{Goldhaber-Gordon
  et~al.}(1998{\natexlab{b}})\citenamefont{Goldhaber-Gordon, G\"ores, Kastner,
  Shtrikman, Mahalu, and Meirav}}]{goldhabergordon1998a}
\bibinfo{author}{\bibfnamefont{D.}~\bibnamefont{Goldhaber-Gordon}},
  \bibinfo{author}{\bibfnamefont{J.}~\bibnamefont{G\"ores}},
  \bibinfo{author}{\bibfnamefont{M.~A.} \bibnamefont{Kastner}},
  \bibinfo{author}{\bibfnamefont{H.}~\bibnamefont{Shtrikman}},
  \bibinfo{author}{\bibfnamefont{D.}~\bibnamefont{Mahalu}}, \bibnamefont{and}
  \bibinfo{author}{\bibfnamefont{U.}~\bibnamefont{Meirav}},
  \bibinfo{journal}{Phys. Rev. Lett.} \textbf{\bibinfo{volume}{81}},
  \bibinfo{pages}{5225} (\bibinfo{year}{1998}{\natexlab{b}}).

\bibitem[{\citenamefont{Parks et~al.}(2010)\citenamefont{Parks, Champagne,
  Costi, Shum, Pasupathy, Neuscamman, Flores-Torres, Cornaglia, Aligia,
  Balseiro et~al.}}]{parks2010}
\bibinfo{author}{\bibfnamefont{J.~J.} \bibnamefont{Parks}},
  \bibinfo{author}{\bibfnamefont{A.~R.} \bibnamefont{Champagne}},
  \bibinfo{author}{\bibfnamefont{T.~A.} \bibnamefont{Costi}},
  \bibinfo{author}{\bibfnamefont{W.~W.} \bibnamefont{Shum}},
  \bibinfo{author}{\bibfnamefont{A.~N.} \bibnamefont{Pasupathy}},
  \bibinfo{author}{\bibfnamefont{E.}~\bibnamefont{Neuscamman}},
  \bibinfo{author}{\bibfnamefont{S.}~\bibnamefont{Flores-Torres}},
  \bibinfo{author}{\bibfnamefont{P.~S.} \bibnamefont{Cornaglia}},
  \bibinfo{author}{\bibfnamefont{A.~A.} \bibnamefont{Aligia}},
  \bibinfo{author}{\bibfnamefont{C.~A.} \bibnamefont{Balseiro}},
  \bibnamefont{et~al.}, \bibinfo{journal}{Science}
  \textbf{\bibinfo{volume}{328}}, \bibinfo{pages}{1370} (\bibinfo{year}{2010}).

\bibitem[{\citenamefont{Sch\"onhammer}(1976)}]{Schonhammer76}
\bibinfo{author}{\bibfnamefont{K.}~\bibnamefont{Sch\"onhammer}},
\bibinfo{journal}{Phys. Rev. B} \textbf{\bibinfo{volume}{13}},
\bibinfo{pages}{4336} (\bibinfo{year}{1976}).
    
\bibitem[{\citenamefont{Gunnarsson and Sch\"onhammer}(1985)}]{Gunnarson85}
\bibinfo{author}{\bibfnamefont{O.}~\bibnamefont{Gunnarsson}}
\bibnamefont{and}
\bibinfo{author}{\bibfnamefont{K.}~\bibnamefont{Sch\"onhammer}},
\bibinfo{journal}{Phys. Rev. B} \textbf{\bibinfo{volume}{31}},
\bibinfo{pages}{4815} (\bibinfo{year}{1985}).
	  

\end{thebibliography}
\end{document}